\documentclass[12pt,preprint]{aastex}

\usepackage{natbib}
\usepackage{ifthen}
\usepackage{amssymb,amsmath}
\usepackage{subfigure}
\newcounter{address}
\newcommand{\latin}[1]{\textit{#1}}

\newcommand{\etal}{\latin{et~al}}
\newlength{\threewidth}
\newlength{\threewidthshort}
\newlength{\twowidth}
\newlength{\twowidthshort}
\newlength{\twothreewidth}
\newlength{\onewidth}
\newcommand{\USNOB}{USNO-B Catalog}
\newcommand{\an}{\textsl{Astrometry.net}}

\newcommand{\numcleantests}{27}
\newcommand{\numnoisytests}{20}

\newcommand{\norm}[1]{\left|\left|#1\right|\right|}
\newcommand{\linespace}{\medskip \\}
\newcommand{\bd}{\textsl{Blind~Date}}
\newcommand{\dchiSq}{\frac{\mathrm{d}\chi^2}{\mathrm{d}t}}
\newcommand{\ddchiSq}{\frac{\mathrm{d}^2\chi^2}{\mathrm{d}t^2}}
\newcommand{\diag}{\mathrm{diag}}
\newcommand{\cutoff}{8}

\newcommand{\examplecaptionA}{Mar-1914 / Nov-1917}
\newcommand{\examplecaptionB}{Mar-1947 / Nov-1945}
\newcommand{\examplecaptionC}{Feb-1949 / May-1951}
\newcommand{\examplecaptionD}{Feb-1914 / Jun-1910}
\newcommand{\examplecaptionE}{Mar-1914 / Nov-1916}
\newcommand{\examplecaptionF}{Jan-1975 / Jan-1800}

\newcommand{\imagex}[1]{x_{#1}}
\newcommand{\imagey}[1]{y_{#1}}
\newcommand{\nextimagex}[1]{{\imagex{#1}}'}
\newcommand{\nextimagey}[1]{{\imagey{#1}}'}
\newcommand{\imagesigmax}[1]{\sigma_{x{#1}}}
\newcommand{\imagesigmay}[1]{\sigma_{y{#1}}}

\newcommand{\catalogx}[1]{u_{#1}}
\newcommand{\catalogy}[1]{v_{#1}}
\newcommand{\catalogsigmax}[1]{ \sigma_{u{#1}} }
\newcommand{\catalogsigmay}[1]{ \sigma_{v{#1}} }

\newcommand{\catalogra}[1]{\ra_{#1}}
\newcommand{\catalogdec}[1]{\dec_{#1}}
\newcommand{\catalogsigmara}[1]{\sigma_{\ra{#1}}}
\newcommand{\catalogsigmadec}[1]{\sigma_{\dec {#1}}}

\newcommand{\catalogmura}[1]{\hat{\mu}_{\ra {#1}}}
\newcommand{\catalogmudec}[1]{\mu_{\dec {#1}}}
\newcommand{\catalogsigmamura}[1]{\hat{\sigma}_{\mu \ra {#1}}}
\newcommand{\catalogsigmamudec}[1]{\sigma_{\mu \dec {#1}}}

\newcommand{\catalogmuraRAW}[1]{\mu_{\ra {#1}}}
\newcommand{\catalogsigmamuraRAW}[1]{\sigma_{\mu \ra {#1}}}
\newcommand{\tearlysym}{t_{\mathrm{early}}}
\newcommand{\tlatesym}{t_{\mathrm{late}}}
\newcommand{\tearly}{1955.0}
\newcommand{\tlate}{1990.0}
\newcommand{\epoch}{2000.0}

\newcommand{\pairdE}{d}
\newcommand{\paird}[1]{\pairdE_{#1}}
\newcommand{\nextpaird}[1]{{\paird{#1}}'}
\newcommand{\pairsigmaE}{\bar{\sigma}}
\newcommand{\pairsigma}[1]{\pairsigmaE_{#1}}
\newcommand{\pairweight}[1]{w_{#1}}

\newcommand{\imageresolution}{\theta_\mathrm{pix}}
\newcommand{\xytord}{WCS_{xy \rightarrow \mathrm{RD}}}
\newcommand{\rdtoxy}{WCS_{\mathrm{RD} \rightarrow xy}}
\newcommand{\ra}{\mathrm{RA}}
\newcommand{\dec}{\mathrm{Dec}}
\newcommand{\unitf}[1]{\mathrm{#1}}
\renewcommand{\arcsec}{\unitf{arcsec}}
\newcommand{\pix}{\unitf{pix}}
\newcommand{\yr}{\unitf{yr}}

\begin{document}
\title{
  \textsl{Blind Date:}\ 
  Using proper motions to determine the ages of historical images
}

\author{
  Jonathan~T.~Barron\altaffilmark{\ref{Toronto},\ref{NYUCS}},
  David~W.~Hogg\altaffilmark{\ref{NYUCCPP},\ref{email}},
  Dustin~Lang\altaffilmark{\ref{Toronto}},
  Sam~Roweis\altaffilmark{\ref{Toronto},\ref{Google}}
}

\setcounter{address}{1}
\altaffiltext{\theaddress}{\stepcounter{address}\label{Toronto}
Department of Computer Science, University of Toronto, 6 King's
College Road, Toronto, Ontario, M5S~3G4 Canada}
\altaffiltext{\theaddress}{\stepcounter{address}\label{NYUCS}
Department of Computer Science, The Courant Institute of Mathematical
Sciences, New York University, 715 Broadway, New York, NY 10003}
\altaffiltext{\theaddress}{\stepcounter{address}\label{NYUCCPP} Center
for Cosmology and Particle Physics, Department of Physics, New York
University, 4 Washington Place, New York, NY 10003}
\altaffiltext{\theaddress}{\stepcounter{address}\label{email} To whom
correspondence should be addressed: \texttt{david.hogg@nyu.edu}}
\altaffiltext{\theaddress}{\stepcounter{address}\label{Google} Google Inc,
Mountain View, CA}

\begin{abstract}
Astrometric calibration is based on patterns of cataloged stars and
therefore effectively assumes a particular epoch, which can be
substantially incorrect for historical images. With the known proper
motions of stars we can ``run back the clock'' to an approximation of
the night sky in any given year, and in principle the year that best
fits stellar patterns in any given image is an estimate of the year in
which that image was taken. In this paper we use 47 scanned
photographic images of M44 spanning years 1910--1975 to demonstrate
this technique. We use only the pixel information in the images; we
use no prior information or meta-data about image pointing, scale,
orientation, or date. \bd\ returns date meta-data for the input
images.  It also improves the astrometric calibration of the image
because the final astrometric calibration is performed at the
appropriate epoch. The accuracy and reliability of \bd\ are functions
of image size, pointing, angular resolution, and depth; performance is
related to the sum of proper-motion signal-to-noise ratios for catalog
stars measured in the input image.  All of the science-quality images
and $85$~percent of the low-quality images in our sample of
photographic plate images of M44 have their dates reliably determined
to within a decade, many to within months.
\end{abstract}


\section{Introduction}

Astronomy needs well calibrated data to make precise measurements, but also
wants to make use of large data sources that are poorly calibrated. Unreliable
data sets such as historical archives, amateurs collections, and engineering
data contain important information, especially in the time domain. Astronomy
needs methods by which data of unknown provenance, quality, and calibration
can be vetted, calibrated, and made reliably useable by the community.

There is an enormous amount of information about proper motions,
stellar and AGN variability, transients, and Solar System bodies in
historical plate archives and the collections of good amateur
astronomers. The Harvard College Observatory Astronomical Plate
Stacks\footnote{http://tdc-www.harvard.edu/plates/} alone contain
enough photographic exposures to cover the entire sky 500 times over,
and span many decades with good coverage and imaging depth. However,
in many cases, it is challenging to use those images
quantitatively. Often the details of observing date, telescope
pointing, bandpass, and exposure time are lost because the logs have
been lost, because the information was written incorrectly or
illegibly, or because it is difficult or expensive to associate each
image with the appropriate record in the log.

The astronomical world is moving towards the development of a Virtual
Observatory, in which a heterogeneous set of data providers
communicate with researchers and the public through open data-sharing
protocols\footnote{http://www.ivoa.net/}.  These protocols can be
easily spoofed---intentionally and unintentionally---and permit the
dissemination of badly calibrated, erroneous, or untrustworthy
data\footnote{http://www.ivoa.net/Documents/Notes/IVOArch/IVOArch-20040615.html}. Indeed,
the lack of a ``trust'' model may compromise the VO's goals of making
it easy for astronomers to discover and use wide varieties of input
data; without trust, all the time that saved in searching and using
has to be spent in verification and tracking of provenance.

Amateur astronomers and educators at college/planetarium observatories are, in
many cases, well equipped and can take science-quality observations,
especially for time-domain science. In addition, most of these astronomers are
interested in contributing to research astronomy. However, it is challenging
at present for these potential data providers to provide their data to the
community in a form that is trivially useable by the research community. These
observers need automated systems to calibrate their data and hardware and to
package the data and meta-data in standards-compliant forms. Even professional
observatories often produce data with incorrect or standards-violating
meta-data because of telescope faults, software bugs, or the youth of most
standards and conventions for digital data formats.

We have begun a large project (\an) to vet, restore, determine, and package in
standards-compliant form the calibration meta-data for astronomical images for
which such information is lost, damaged, or unreliable \citep{lang08a}. Our
system can astrometrically calibrate an image (that is, determine the world
coordinate system) using the information in the image pixels alone. It starts
by identifying asterisms that determine astrometric calibration. Once the
astrometry is correct, the sources in the image can be identified with
catalogs and other calibration meta-data can be inferred through quantitative
image analysis. In the process of testing and running this calibration system,
we have indeed confirmed that many historical and amateur images---and even
some scientific images from modern professional facilities---have missing or
incorrect astrometric meta-data; automated calibration, as a vetting step at
the very least, is essential for all data sources.

For the same reason that the time domain is interesting, it can also be used
to calibrate the \emph{date} at which an image was taken. Stars are moving and
varying, so the particular configuration and relative brightnesses of the
stars in an image provide, in principle, a measure of the time at which the
image was taken. Here we show that stellar motions catalogued at the present
day can be used to age-date plates from a plate archive to within a few years,
even for very old plates. This new capability takes the \an\ project a small
step closer to being a comprehensive image meta-data vetting and automated
calibration system.

\section{Input data}

The larger goal of our project is to add calibration meta-data to data of
unknown provenance. For this reason, the system begins truly ``blind'' in the
sense that we ignore all meta-data associated with each input image, and start
with only the image pixels themselves. We calibrate the images using the \an\
astrometric system; this calibration provides output that is taken as input
data for the \bd\ analysis.

For each automatically detected star $i$ in the image, there is a centroid
$(\imagex{i}, \imagey{i})$ in the input image measured in pixels. For each
source, this centroid is the location of the maximum of a second-order
polynomial (generalized parabolic) surface fit to the area immediately
surrounding the center of the image star. The fit also provides uncertainties
$(\imagesigmax{i}, \imagesigmay{i})$ in these centroids. For unsaturated stars
these are taken (arbitrarily) to be one pixel, and for saturated stars (which
are common in the digitized plate data) these are taken to be one-third of the
radius of the saturated region. These are over-estimates, since the stars
are detected at good signal-to-noise, but this is
conservative; furthermore, in the case of saturated stars, it is possible for
the saturated ``disk'' to be non-concentric with the true centroid.

The system also provides a rough world coordinate system (WCS) for the
image; that is, first guesses at two functions: $\xytord(x,y)$, which
transforms points from the image plane in pixels into celestial
coordinates in angular units, and $\rdtoxy(\ra,\dec)$, the inverse. A
derived quantity from these functions is the pixel scale
$\imageresolution$, measured in angle per pixel, which we will use
below.  Strictly $\imageresolution$ is a function of position in the
image, but in typical science images it does not vary substantially.

The WCS effectively identifies the sources from the \USNOB\ \citep{monet03a}
that are in or likely to be in the image. For each catalog ``star'' $j$ (the
\USNOB\ contains both stars and compact galaxies) inside the image, the
catalog contains a J2000 celestial position $(\catalogra{j}, \catalogdec{j})$
on the celestial sphere (extrapolated to epoch $\epoch$) measured in angular
units, an uncertainty $(\catalogsigmara{j}, \catalogsigmadec{j})$ in that
position, a proper motion $(\catalogmuraRAW{j}, \catalogmudec{j})$ measured in
angle per time, and an uncertainty
$(\catalogsigmamuraRAW{j},\catalogsigmamudec{j})$ in that proper motion.

The \USNOB\ uncertainties required some processing and adjustment. For the
sake of clarity and simplicity, we make as few assumptions as possible in
transforming the uncertainties. Our approach here is not intended to be
definitive. Many of these entries in the catalog have values of zero for the
uncertainty of position or proper motion. In the case of zero-valued
uncertainty in position, we assume that the uncertainty is below the precision
of $0.002\,\arcsec$ at which the catalog was reported. We therefore set all
zero-valued position uncertainties to one-half of the precision
($0.001\,\arcsec$). For entries with zero-valued uncertainty in proper motion,
there is more we need to consider; a nonzero-valued proper motion paired with
a zero-valued uncertainty indicates that the uncertainty is below the
precision of the catalog, and therefore should be set to half of the
precision. A zero-valued proper motion paired with a zero-valued uncertainty
indicates that the proper motion of the entry could not be measured accurately
(Dave Monet, private communication). We therefore set the uncertainty in the
proper motion for such entries to three times the median value of nonzero
proper motion uncertainties. This captures the idea that, generally speaking,
we are significantly more uncertain about the proper motion of such entries
than we are most other entries. A more principled approach could certainly be
attempted but we found that this worked well enough for our purposes.

Much work has already been done by the \an\ team to identify spurious sources
in the \USNOB\ that appear to have been created by diffraction spikes and
reflection halos \citep{barron08a}. For the purposes of this project, we
ignore the entries in the catalog which have been flagged as spurious. Testing
has shown that ignoring these sources generally improves the accuracy of our
results.

Using a technique that will be described in a future paper from the
\an\ team, we are able to estimate the bandpass of each image being
processed, in that we determine which bandpass of the \USNOB\ most
closely predicts the brightness ordering of the stars in the
image. Though this technique is still in an experimental stage, its
results are not very controversial; all of Harvard's images of M44
appear to best match the blue bands ($O$ and $J$ emulsions) of the
\USNOB. This finding is reinforced through experimentation with
manually setting each image's band: On average, \bd\ performs better
on these images using the $J$ emulsion of the Catalog
than on any other band.

Assuming that we have estimated the bandpass correctly, the $N$ stars in the
image should correspond---roughly---to the $N$ brightest catalog stars that
lie within the area of the image. For that reason, we use only the $N$
brightest catalog stars in what follows.

\section{Method}

We use the \USNOB\ positions and proper motions to ``wind'' the $N$ catalog
stars backwards and forward through time along the celestial sphere. Once the
catalog has been adjusted, we can use the image WCS to project the catalog
entries onto the image plane, making a synthetic catalog for that image at
that time, in image coordinates. We then attempt to fit the image stars to
that synthetic image of the moved catalog stars. We choose a freedom with
which the image star positions are allowed to warp to fit the catalog star
positions, and a scalar objective function that is minimized when the
positions are ``best'' warped. We warp the input image to the catalog
``wound'' to different times, and use the best-fit values of the objective
function to determine the year at which the image was taken.

\subsection{Winding back the catalog}

We estimate the celestial coordinates of catalog star $j$ at the arbitrary
date $t$, and then project them onto the image plane
\begin{eqnarray}\displaystyle
	\left( \catalogx{j}, \catalogy{j} \right) & = &
	\rdtoxy \left(
	\catalogra{j} - \catalogmura{j} [t-(\epoch\,\yr)]
	,
	\catalogdec{j} - \catalogmudec{j} [t-(\epoch\,\yr)]
	\right)
	\quad ,
\end{eqnarray}
where we have adjusted the $\catalogmuraRAW{j}$ proper motion vector
components and their associated uncertainties into coordinate derivatives
$\catalogmura{j}$ by
\begin{eqnarray}\displaystyle
\catalogmura{j} & = & \mathrm{cos}(\catalogdec{j})\,\catalogmuraRAW{j}
\nonumber \\
\catalogsigmamura{j} & = & \mathrm{cos}(\catalogdec{j})\,\catalogsigmamuraRAW{j}
\quad .
\end{eqnarray}
We estimate the uncertainty of the location of catalog star $j$ at year $t$ on
the image plane with
\begin{eqnarray}\displaystyle
	\catalogsigmax{j}
	& = &
	\frac{1}{\imageresolution}\,\sqrt{\catalogsigmara{j}^{2} + ( \max\left( | t - \tearlysym |, | t - \tlatesym | \right)\,\catalogsigmamura{j} )^{2}}
	\nonumber \\
	\catalogsigmay{j}
	& = &
	\frac{1}{\imageresolution} \sqrt{\catalogsigmadec{j}^{2} + ( \max\left( | t - \tearlysym |, | t - \tlatesym | \right)\,\catalogsigmamudec{j} )^{2}}
\label{eq:uncertainty}
\end{eqnarray}
where $\tearlysym$ and $\tlatesym$ are the dates at which the \USNOB\ source
imagery were taken, which in this patch of the sky are \tearly\ and \tlate,
respectively. All positions in the Catalog, however, were extrapolated to the
year $\epoch$ (epoch and equinox). This means that though we are given each
catalog star's location at the year $\epoch$, we know that each star's
measured location is, in general, more accurate between the two epochs in
which the images were taken, and less accurate at years further from that
range. This means that when ``winding'' locations through time, we look at the
difference between $t$ and the year $\epoch$; when ``winding'' uncertainties
through time, we look at the maximum distance between $t$ and both
$\tearlysym$ and $\tlatesym$. This is equivalent to using
$(\tearlysym + \tlatesym)/2$ as our reference year, and assuming a non-zero
uncertainty on the measurement of each star's proper motion at that reference
year.

Note that in our notation for $(\catalogx{j}, \catalogy{j})$ and
$(\catalogsigmax{j}, \catalogsigmay{j})$, we do not reference $t$. This is
because once the catalog has been wound through time and projected onto the
image plane, we consider time to be fixed. Note that in Sections
\ref{sec:objectiveSection} and \ref{sec:fittingSection}, time will remain
fixed, and therefore $t$ is not mentioned in any of the notation except for in
$\chi^2(t)$.

\subsection{Objective function}
\label{sec:objectiveSection}

Determination of the image coordinate system and date involves finding
parameters---astrometric parameters and the date---that optimize an objective
function. The choice of this objective is therefore the fundamental scientific
choice in the project.

We seek an objective function that has the following properties, listed in
rough order of priority: The function must decrease as image-coordinate
distances between catalog and image stars decrease. The function must be
insensitive to anomalous outliers, and more sensitive to well-measured stars
than to poorly-measured stars. The function must be some approximation to a
likelihood or have some equivalent justification so that changes in the
function with respect to parameters can be interpreted in terms of
uncertainties in those parameters. The function ought to be differentiable and
second-differentiable with respect to all fit parameters (in particular time
and astrometric calibration). The function should be easily optimized. We have
identified an objective function that has all of these properties; it is so
similar to the least-square function that we call it a ``modified
chi-squared'' and denote it ``$\chi^2$''.

For all $i = 1:N$ and $j = 1:N$ we compute $\paird{ij}$, the Euclidian
distance between image star $i$ and catalog stars $j$ in the image plane.
\begin{equation} \paird{ij} = \sqrt{ \left( \catalogx{j} - \imagex{i}
\right)^2 + \left( \catalogy{j} - \imagey{i} \right)^2 } \end{equation}

\begin{figure}[t]

	\centering
	\resizebox{\twowidthshort}{!}{\includegraphics{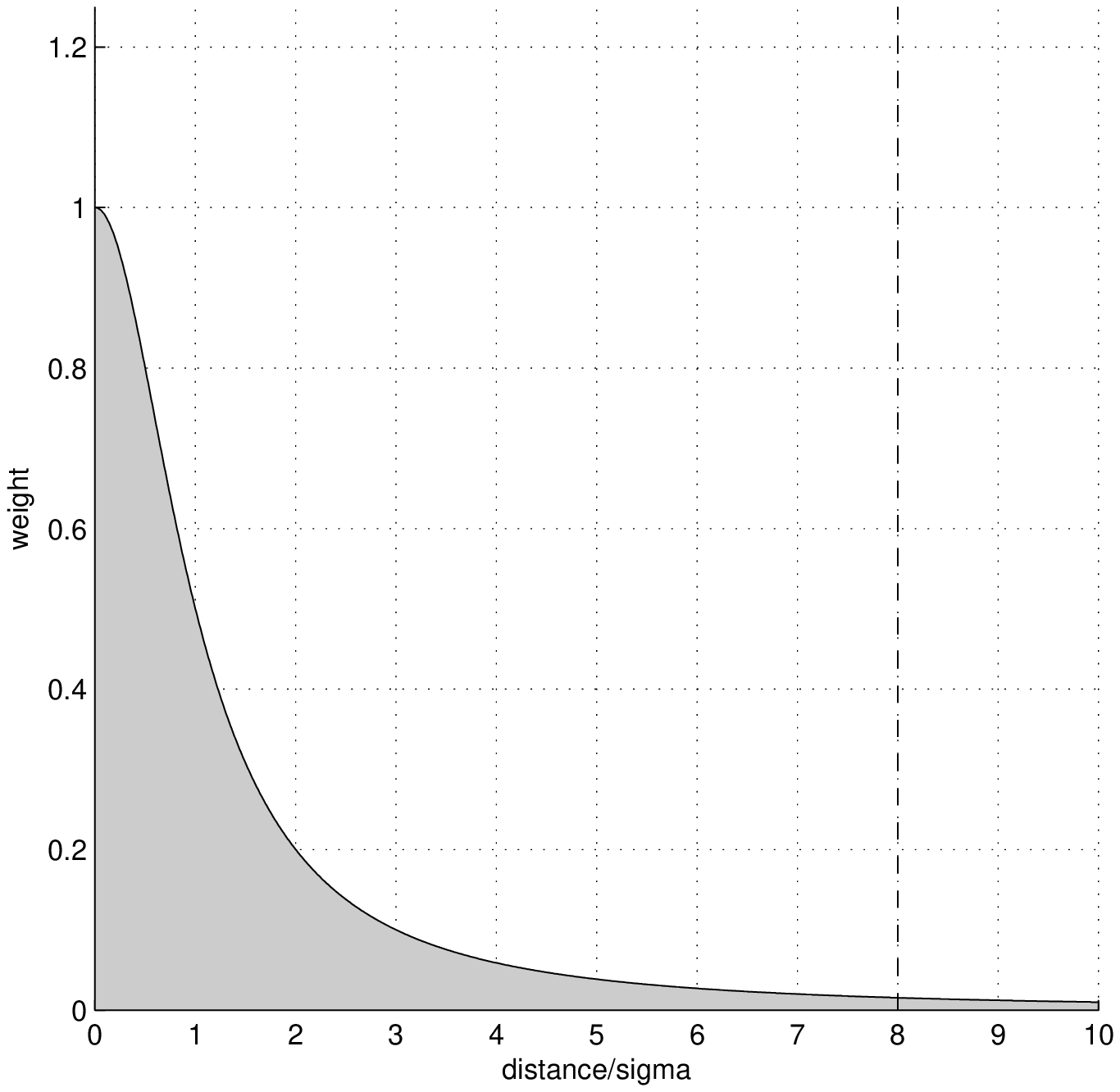}}
	\resizebox{\twowidthshort}{!}{\includegraphics{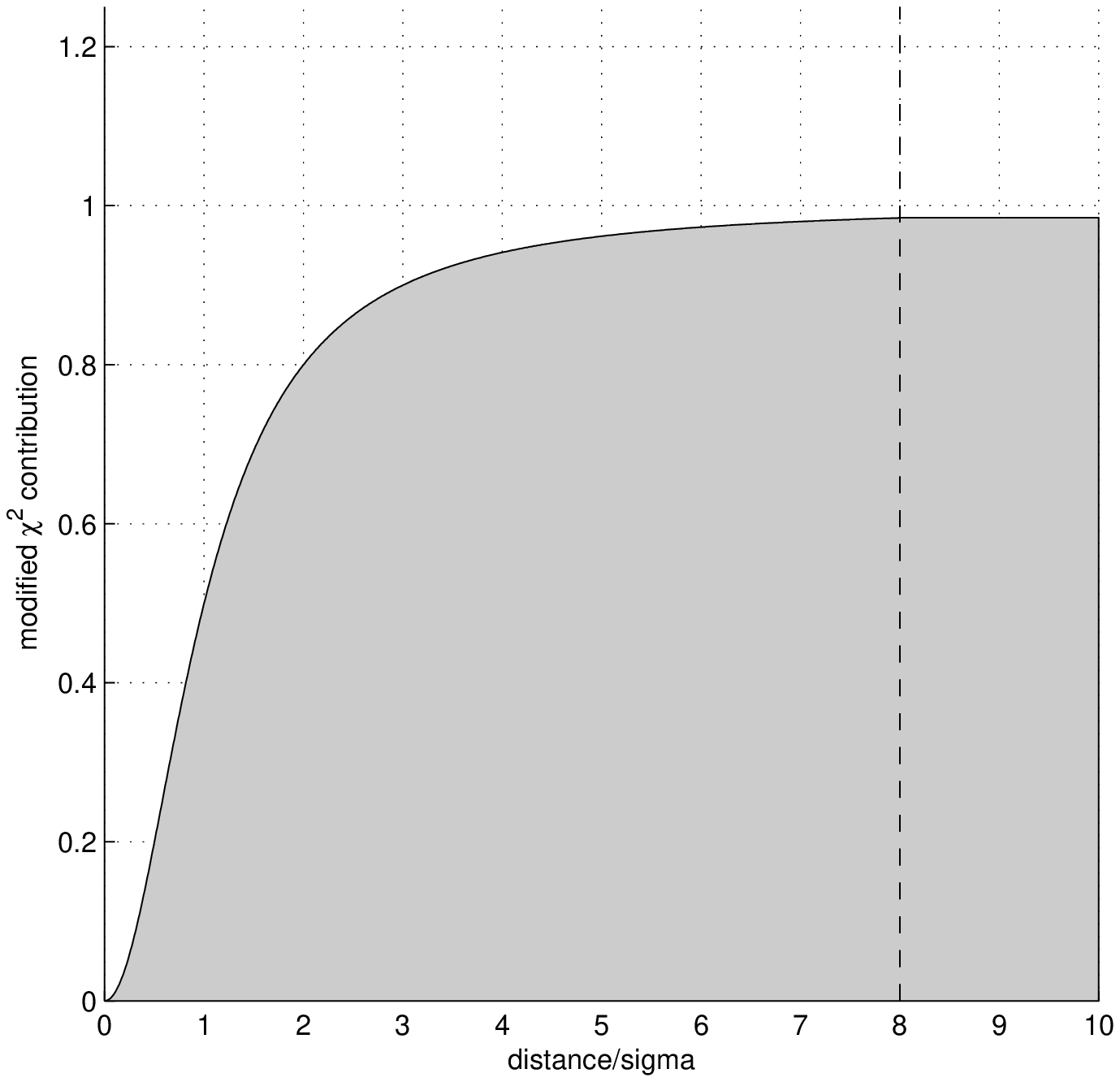}}

	\caption{On the left, the weighting function versus a pair's ratio of
	distance to uncertainty. On the right, the corresponding weighted
	contribution of a pair to the modified chi-squared. The dotted line
	indicates the point at which we assume that the weighted contribution
	stops changing.
	\label{fig:weightCurve}}
\end{figure}

We also estimate the uncertainty $\pairsigma{ij}$ of each pair's distance
measurement, using the previously defined values for
$(\imagesigmax{i}, \imagesigmay{i})$ and
$(\catalogsigmax{j}, \catalogsigmay{j})$. We calculate
the combined uncertainty for the pair in $x$ and $y$ at time $t$, and then
simply take the mean as an estimation of the uncertainty of that pair:

\begin{equation}
\pairsigma{ij} = 
\frac{1}{2}\,\left(\sqrt{ \imagesigmax{i}^2 + \catalogsigmax{j}^2}
  + \sqrt{ \imagesigmay{i}^2 + \catalogsigmay{j}^2}\right)
\end{equation}
We define a weighting function for each pair that returns a value between 0
and 1 based on the ratio of the distance of a pair to the uncertainty of that
pair:
\begin{equation}
W(\pairdE, \pairsigmaE) = \frac{1}{1+\left(\frac{\pairdE}{\pairsigmaE}\right)^2}
\end{equation}
(see also Figure~\ref{fig:weightCurve}).
This function has the property that outliers are down-weighed in a smooth
manner; it causes the influence of a image--catalog pair to smoothly drop to
zero at large displacement. This permits us to avoid the
discontinuous optimization problem of sigma clipping, which is the standard
``robust estimation'' technique in common use in astronomy applications. The
weighting function has a number of properties which make it well-suited to our
purpose:
\begin{eqnarray}\displaystyle
  W(0, \pairsigmaE) & = & 1
  \label{eqn:wnearzero} \\
  \left.\frac{\mathrm{d}W}{\mathrm{d}\pairdE}(\pairdE, \pairsigmaE)\right|_{d=0} & = & 0
  \label{eqn:dwnearzero} \\
  \lim_{(\pairdE/\pairsigmaE)\rightarrow\infty} W(\pairdE, \pairsigmaE)\left(\frac{\pairdE}{\pairsigmaE}\right)^2 & = & 1 
  \label{eqn:wbig}
\end{eqnarray}

We use this weighting function to assign a weight $\pairweight{ij}$ to all
pairs as follows:
\begin{equation}
\pairweight{ij} = W \left(\paird{ij}, \pairsigma{ij} \right)
\end{equation}
Our final objective function is:
\begin{equation}
\displaystyle \chi^2(t) =
	\sum_{ij} \pairweight{ij}\,\left(\frac{\paird{ij}}{\pairsigma{ij}}\right)^2
\end{equation}
where the sum is over all possible image--catalog pairs.

For ``good'' (small-separation) image--catalog pairs, the weight function is
near unity (equation \ref{eqn:wnearzero}) and has near-zero derivatives
(equation \ref{eqn:dwnearzero}), so small changes in separation do not enter
strongly into derivatives of the objective function. For ``bad''
(large-separation) image--catalog pairs, the pair's contribution to
$\chi^2(t)$ is nearly constant (equation \ref{eqn:wbig}). This makes
optimization and interpretation of our objective function very like
optimization and interpretation of a chi-squared fitting system. This weighted
chi-squared objective function could also be interpreted as a Geman-McLure
error function, In this framework, optimizing the objective function is
equivalent to robust M-estimation \citep{hampel86}.

In constructing the modified chi-squared, we use---in principle---all
image--catalog pairs, irrespective of their separation in image coordinates.
The fact that the contribution of a pair to the objective function quickly
converges to $1$ as the pair becomes highly separated allows us---in
practice---to ignore all highly separated pairs. We therefore choose to
approximate the contributions of all pairs where
$\pairdE > \cutoff\,\pairsigmaE$ as
$W(\cutoff\,\pairsigmaE, \pairsigmaE) \left( \frac{8\pairsigmaE}{\pairsigmaE} \right)^2 $,
which is $64/65$. This dramatically speeds up our computation.

\subsection{Fitting the image}
\label{sec:fittingSection}

At this point, we have our image stars on the image plane, our catalog stars
(wound to the time of interest) projected onto the image plane, and an
objective function that we wish to minimize. We need to find the
transformation that we can apply to the image that minimizes the objective
function, which we take to be the transformation that best ``fits'' the image
to the catalog.

Testing has suggested that the initial location of the image returned by the
\an\ solver is close enough to the optimal location that locally minimizing
the objective function is sufficient for finding the global minimum, and that
we generally do not run the risk of falling into a false local minimum.
Therefore, we only present our method for locally minimizing the optimal
function through iteratively reweighted least-squares (IRLS). We experimented
with techniques such as RANSAC to fit images to the catalog in the face of
extreme noise, but no technique was more effective and robust than our IRLS
method.

Let us first construct a solution to a simplified version of this problem, in
which we know which correspondences are true: We assume that image point $i$
corresponds to catalog point $i$ for all $i \in \{ 1, 2, ..., N \} $. Assuming
that we are interested in solving for an affine transformation (first-order
linear transformation plus shift), this means that we need to find the
transformation matrix that best satisfies the following equations:
\begin{equation}
	\underset{i \in \{ 1, 2, ..., N \} }{\forall} \,
	\left[\begin{array}{ccc}
	m_x & m_y & t_x \\
	n_x & n_y & t_y
	\end{array} \right]
	\left[\begin{array}{c}
	\imagex{i} \\
	\imagey{i} \\
	1
	\end{array} \right]
	=
	\left[\begin{array}{c}
	\catalogx{i} \\
	\catalogy{i}
	\end{array} \right]
\end{equation}

This can be generalized straightforwardly for higher order transformations.

The transformation that best satisfies these equations can be found using a
standard least-squares solver. We can then use this transformation to warp all
of the image points onto the catalog points (and vice-versa), thus solving our
simplified problem.

Of course, since we do not know which image stars correspond to which catalog
stars, we must include equations for all image--catalog pairs. We are not
interested in the solution to this problem, as it would describe a
transformation from \emph{every} image star to \emph{every} catalog star. To
specify a transformation that satisfies \emph{likely} image--catalog
correspondences, we must use our weighting function to make soft assignments
regarding correspondences. We therefore use the following equations:
\begin{equation}
\underset{i \in \{1, 2, ..., N \} }{\forall} \,\,\,
\underset{j \in \{1, 2, ..., N \} }{\forall} \,
\left[\begin{array}{cc}
\frac{\sqrt{\pairweight{ij}}}{\pairsigma{ij}} & 0 \\
0 & \frac{\sqrt{\pairweight{ij}}}{\pairsigma{ij}}
\end{array} \right]
	\left[\begin{array}{ccc}
	m_x & m_y & t_x \\
	n_x & n_y & t_y
	\end{array} \right]
	\left[\begin{array}{c}
	\imagex{i} \\
	\imagey{i} \\
	1
	\end{array} \right]
	=
	\left[\begin{array}{cc}
	\frac{\sqrt{\pairweight{ij}}}{\pairsigma{ij}} & 0 \\
	0 & \frac{\sqrt{\pairweight{ij}}}{\pairsigma{ij}}
	\end{array} \right]
	\left[\begin{array}{c}
	\catalogx{j} \\
	\catalogy{j}
	\end{array} \right]
\end{equation}

We begin our solution by constructing a linear system which contains all of
the previously described (unweighted) equations.
\begin{equation}
	\left[\begin{array}{cccccc}
	\imagex{1} & \imagey{1} & 1 & 0 & 0 & 0\\
	0 & 0 & 0 & \imagex{1} & \imagey{1} & 1\\
	\imagex{1} & \imagey{1}  & 1 & 0 & 0 & 0\\
	0 & 0 & 0 & \imagex{1} & \imagey{1} & 1\\
	 & & \ldots \\
	\imagex{N} & \imagey{N} & 1 & 0 & 0 & 0\\
	0 & 0 & 0 & \imagex{N} & \imagey{N} & 1\\
	\imagex{N} & \imagey{N} & 1 & 0 & 0 & 0\\
	0 & 0 & 0 & \imagex{N} & \imagey{N} & 1
	\end{array} \right]		
	\left[\begin{array}{c}
	m_x \\
	m_y \\
	t_x \\
	n_x \\
	n_y \\
	t_y 
	\end{array} \right]
	=
	\left[\begin{array}{c}
	\catalogx{1} \\
	\catalogy{1} \\
	\catalogx{2} \\
	\catalogy{2} \\
	\ldots \\
	\catalogx{N-1} \\
	\catalogy{N-1} \\
	\catalogx{N} \\
	\catalogy{N}
	\end{array} \right]
\end{equation}

We can write this matrix equation as:
\begin{equation}
 \mathbf{Ax} = \mathbf{b}
\end{equation}

To introduce the weight values described in the equations, we construct our
weight matrix $ \mathbf{W} $, as follows:
\begin{equation}
\mathbf{W} = \diag
\left(\begin{array}{ccccccccc}
\frac{\sqrt{\pairweight{1,1}}}{\pairsigma{1,1}},
\frac{\sqrt{\pairweight{1,1}}}{\pairsigma{1,1}},
\frac{\sqrt{\pairweight{1,2}}}{\pairsigma{1,2}},
\frac{\sqrt{\pairweight{1,2}}}{\pairsigma{1,2}},
\ldots,
\frac{\sqrt{\pairweight{N,N}}}{\pairsigma{N,N}},
\frac{\sqrt{\pairweight{N,N}}}{\pairsigma{N,N}}
\end{array} \right)
\end{equation}

Our final matrix equation can then be written as:
\begin{equation}
 \mathbf{WAx} = \mathbf{Wb} 
\end{equation}
We then find $\mathbf{x}$ such that the squared residuals,
$\mathbf{ (WAx-Wb)^T (WAx-Wb) }$,
are minimized. By construction, $\mathbf{x}$ describes the transformation that
best satisfies all of the equations we previously constructed, and can be
found using a standard least-squares solver.

Using the transformation defined by $\mathbf{x}$ we can calculate the
coordinates of our warped image points.
\begin{eqnarray}
	\displaystyle
	{\imagex{i}}' & = & m_x \imagex{i} + m_y \imagey{i} + t_x \nonumber \\
	{\imagey{i}}' & = & n_x \imagex{i} + n_y \imagey{i} + t_y
\end{eqnarray}

Unlike the simplified version of this problem, the warp found after one
iteration is not our final solution. As we warp the image, the distances and
weights between the image and catalog points change, and so our objective
function is no longer minimized. The solution is to repeatedly recalculate our
$\paird{ij}$ and $\pairweight{ij}$ values and our weighted least squares
transformation. We perform this iteratively reweighted least-squares operation
until the resulting transformations stop changing, which by construction is
also when our objective function stops changing. The solution which we
converge upon is returned as the best fit of the image onto the catalog.

The IRLS method minimizes the objective function, as it was constructed such
that the objective function is equal to the sum of the squares of the weighted
residuals of the matrix equation.
\begin{eqnarray}\displaystyle
\mathbf{ (WAx-Wb)^T (WAx-Wb) } & = & \mathbf{ (W(Ax-b))^T (W(Ax-b)) } \nonumber \\
& = & \sum_{ij}
\left(\frac{\sqrt{\pairweight{ij}}}{\pairsigma{ij} }
\left( \catalogx{j} - \nextimagex{i} \right) \right)^2 + 
\left(\frac{\sqrt{\pairweight{ij}}}{\pairsigma{ij} }
\left( \catalogy{j} - \nextimagey{i} \right) \right)^2 \nonumber \\
& = & \sum_{ij} \left(\frac{\sqrt{\pairweight{ij}}}{\pairsigma{ij}} \nextpaird{ij} \right)^2 \nonumber \\
& = & \sum_{ij} \pairweight{ij} \left(\frac{{\nextpaird{ij}}}{\pairsigma{ij}}\right)^2 \nonumber \\
& \approx & \chi^2(t) 
\end{eqnarray}

We say that the residuals only \emph{approximate} the optimal function because
we use the weights of the previous transformed image ($\pairweight{ij}$) and
the distances of the next warped transformed image ($\nextpaird{ij}$). Though
this may seem strange, once the IRLS has converged on a solution and the
optimal function has stopped changing, the distances in one iteration are
effectively equal to the distances in the following iteration.

For numerical stability we always use the initial coordinates when
constructing our transformation, and only return the final warped image
coordinates. The intermediate warped image coordinates are only used for
calculating distances and weights. This means that our final image coordinates
have only been subjected to one transformation, as opposed to many small
transformations.

\begin{figure}
\centering
\subfigure[Catalog at year 1914, initial image.]{
	\label{fig:fitExample1}
	\resizebox{\twowidthshort}{!}{\includegraphics{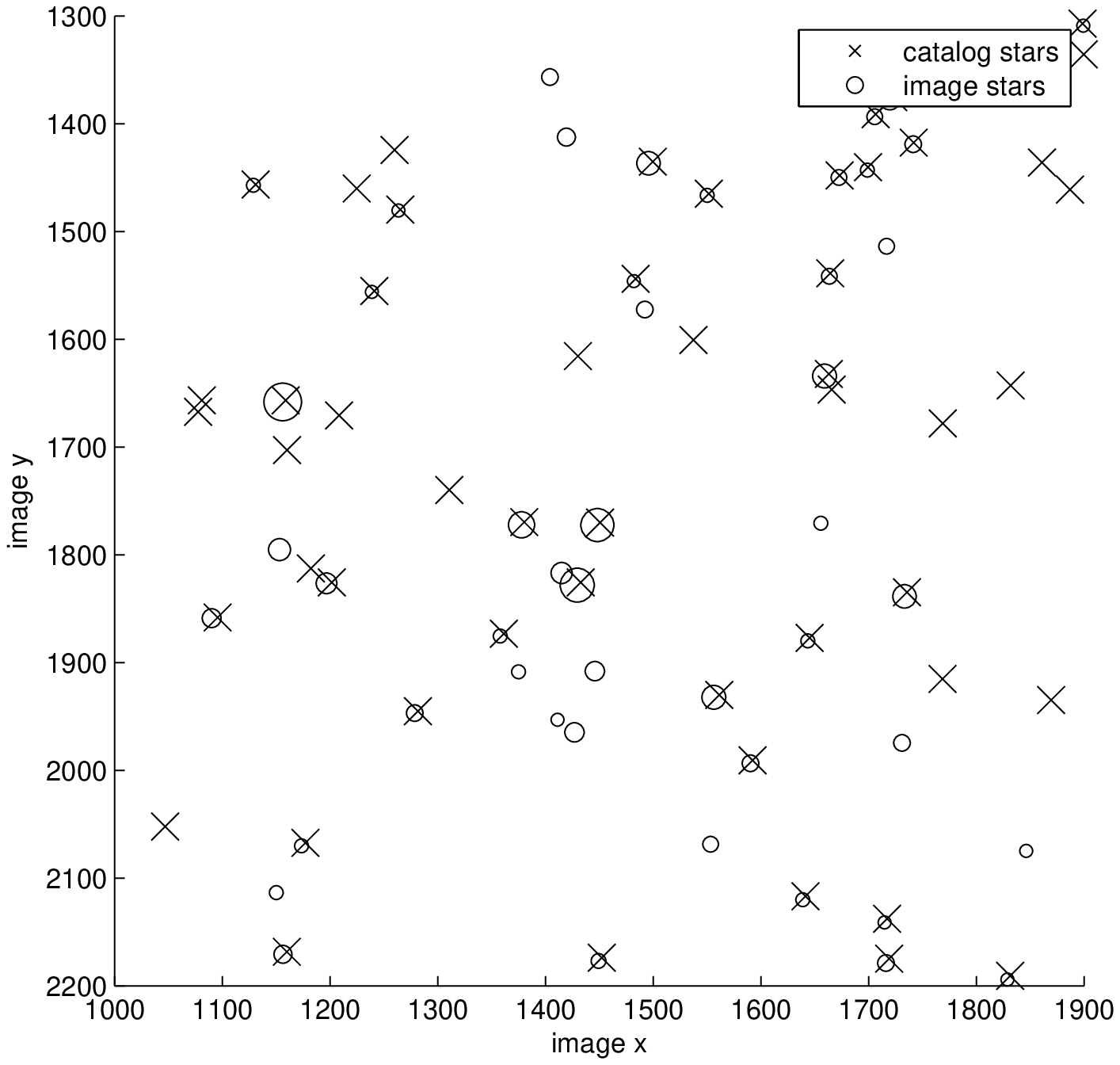}}
}
\subfigure[Catalog at year 1914, fitted image.]{
	\label{fig:fitExample2}
	\resizebox{\twowidthshort}{!}{\includegraphics{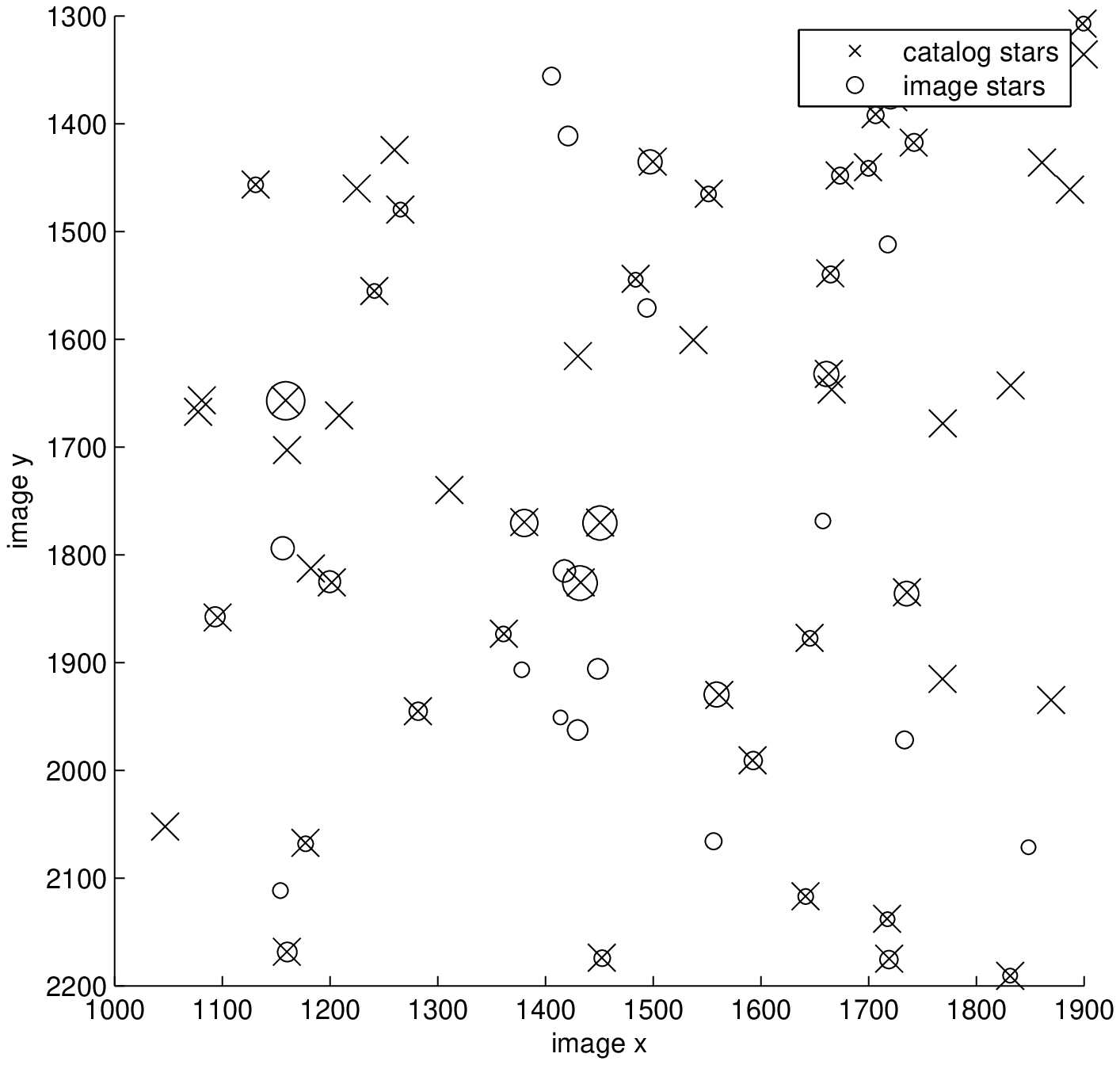}}
}
\subfigure[Catalog at year 2000, initial image.]{
	\label{fig:fitExample3}
	\resizebox{\twowidthshort}{!}{\includegraphics{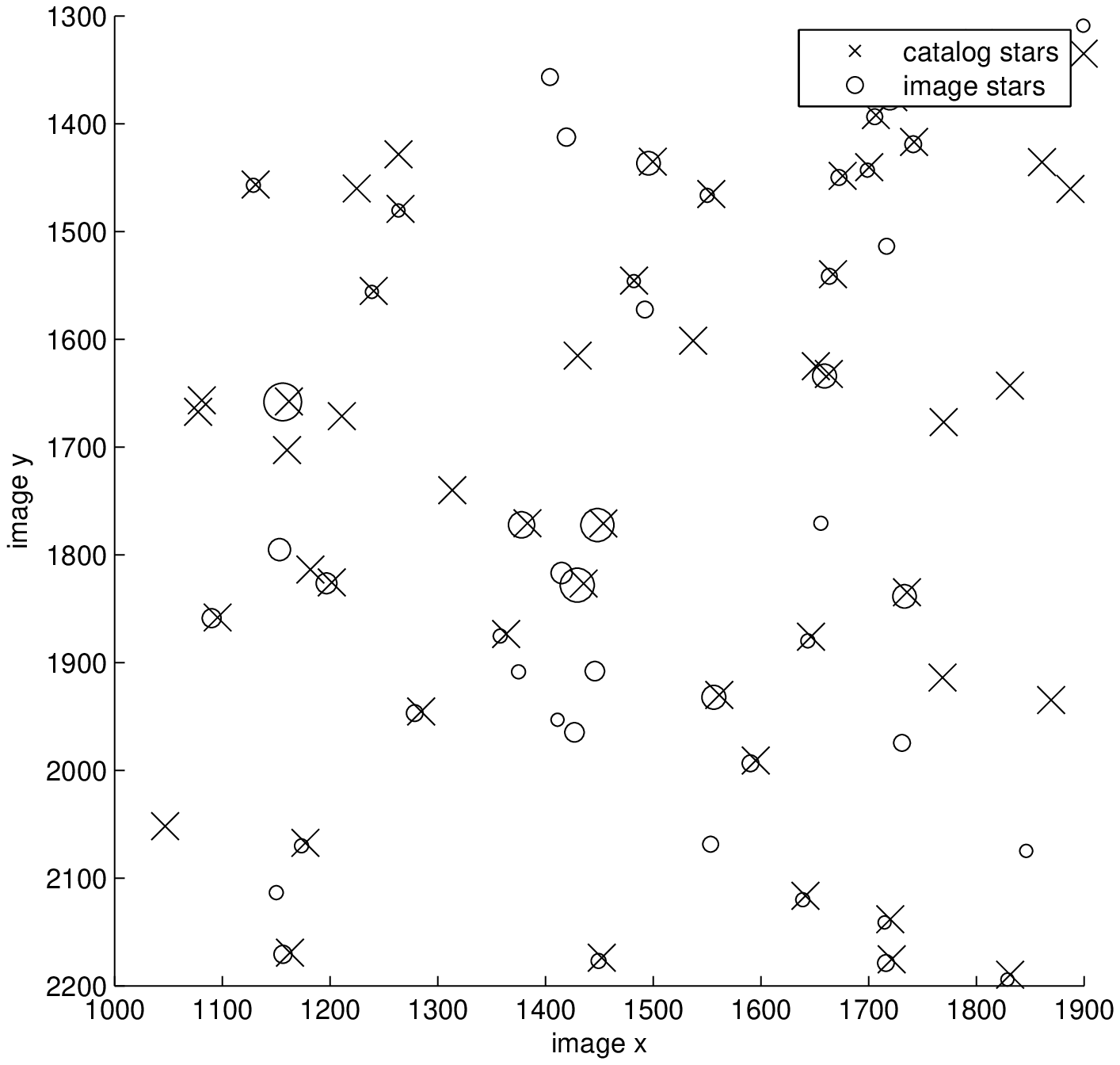}}
}
\subfigure[Catalog at year 2000, fitted image.]{
	\label{fig:fitExample4}
	\resizebox{\twowidthshort}{!}{\includegraphics{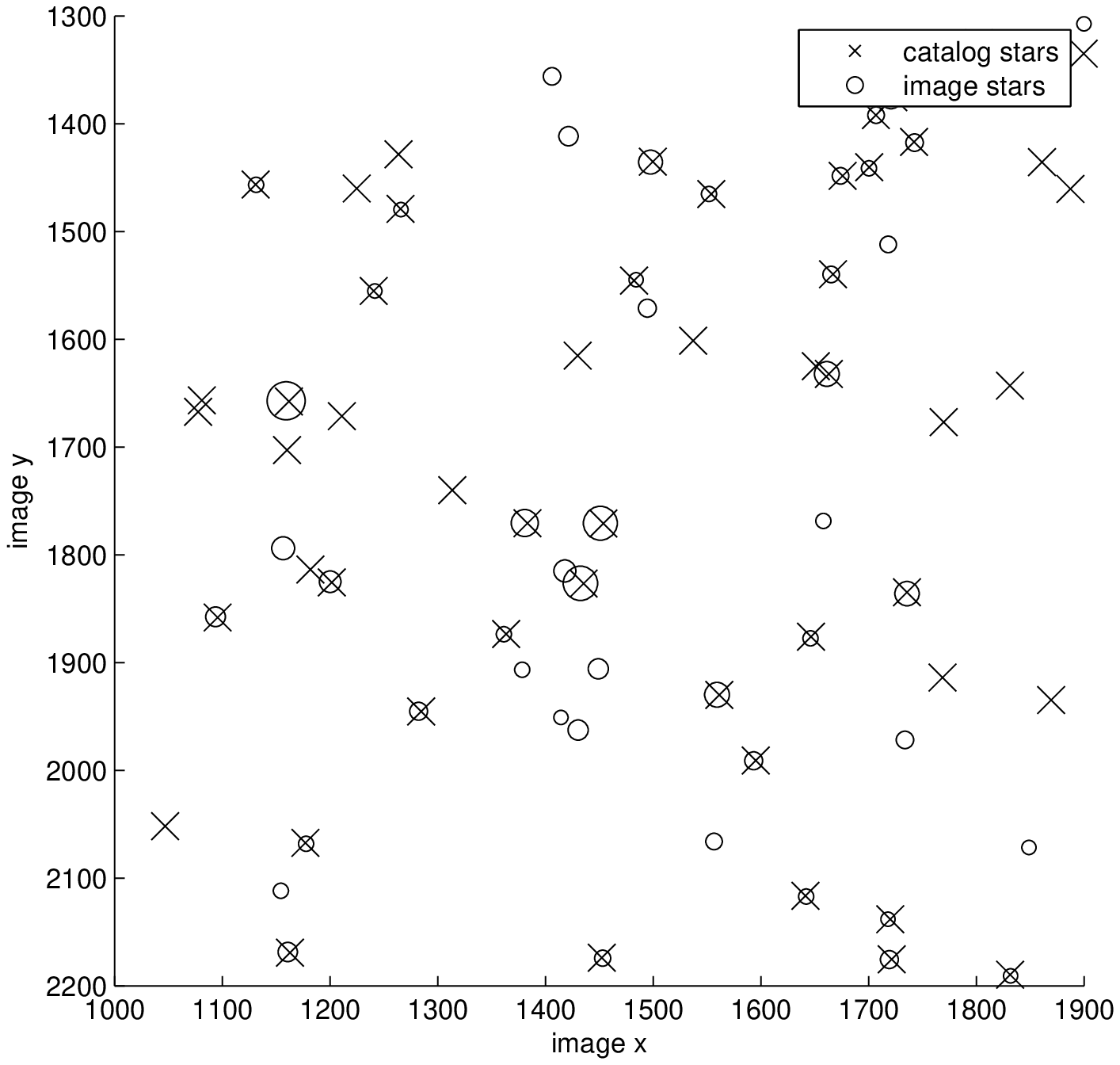}}
}
\caption{
	The extracted sources (size is proportional to brightness) from the 900 by
900 pixel sub-image shown in Figure~\ref{fig:imageExamples}. These plots
illustrate the error in the initial calibration returned by the \an\
solver, as well as the difference in fitting a historical image to \USNOB\ at
the year 2000 and to the Catalog at 1914, the year which we correctly estimate
to be the image's year of origin.
\label{fig:fitExamples}}
\end{figure}

See Figure~\ref{fig:fitExamples} for some examples of fit and unfit images
against correct and incorrect catalog dates.

\subsection{Estimating the date}
\label{sec:searchSection}

Now that we have defined a method for fitting an image to the catalog, and a
metric by which we can assess the degree to which an image can be fit to the
catalog, we can use a number of techniques to estimate the year in which the
image can best be fit to the catalog. We take that year to be the year in
which the image was created.

This problem can be phrased as such: an image has some unknown chi-squared
curve $\chi^2(t)$, of which we wish to find the year $t_0$, the theoretical
optimal year such that: \begin{equation} t_0 = \underset{t}{\arg\min} \,
\chi^2(t) \end{equation}

In the algorithm we will describe, finding $t_0$ is not computationally
feasible, so we must settle on finding a reasonable approximation of the
minimum $t^*$, such that given an accuracy threshold $\delta_\chi$ we are
confident that:
\begin{equation}
\norm{\chi^2(t_0)-\chi^2(t^*)} \leq \delta_\chi
\end{equation}

Once we have found $t^*$ and therefore $\chi^2(t^*)$, we also wish to find the
extents of our uncertainty region, $(t^*_-, t^*_+)$, such that:
\begin{equation}
	\begin{array}{c}
		t^*_- < t^* < t^*_+ \linespace
		\chi^2(t^*_-) = \chi^2(t^*)+1 = \chi^2(t^*_+) \linespace
		\underset{t \in [t^*_-,t^*_+]}{\forall} {\, \chi^2(t) \leq \chi^2(t^*)+1 }
	\end{array}
\end{equation}

Ideally, we want to find all of these values as accurately as possible while
sampling the $\chi^2$ curve as few times as possible. There are a number of
methods by which we can accomplish this task, each with different tradeoffs
concerning efficiency and assumptions about the shape of the curve.

The simplest method for estimating the origin date is through brute force. We
sample our $\chi^2$ curve at regular intervals, and take the year in which our
objective function scored the lowest as $t^*$. We then linearly interpolate
along the curve to find the extents of the uncertainty region. This method is
terribly inefficient and assumes nothing about the shape of the curve, so we
only use it as an approximate ground truth to which we will compare our final
algorithm.

Our search algorithm begins with first sampling our $\chi^2$ curve at a very
broad, regular interval. Though only one initial sample is required, for the
figures shown in this paper we sample the curve at $1900$, $1950$, and $2000$.
We take the sampled year with the lowest $\chi^2$ score to be $t_n$, and we
then iteratively refine $t_n$ into $t_{n+1}$ until we believe that we have
found $t^*$.

Our objective function was constructed such that we could efficiently
calculate $\dchiSq(t)$ and $\ddchiSq(t)$. The equations for these analytical
derivatives are elaborate, so we do not present them here. Since $\chi^2(t^*)$
is a minimum in the $\chi^2$ curve, we can assume that $\dchiSq(t^*)=0$. We
can therefore use Newton's method to find $t_{n+1}$:
\begin{equation}
	t_{n+1} = t_n - \frac{\dchiSq(t_n)}{\ddchiSq(t_n)}
\end{equation}

We iteratively evaluate $t_{n+1}$ until
$\norm{\chi^2(t_{n+1})-\chi^2(t_n)} \leq \delta_\chi$
holds for two consecutive iterations, at which point we take $t_{n+1}$ as
$t^*$. Since Newton's method generally converges quadratically, this is an
extremely fast process.\footnote{Of course since $\chi^2(t)$ is not globally
quadratic, Newton's method is not guaranteed to converge from any
starting point. To improve numerical conditioning and ensure that 
the search remains well-behaved in the face of somewhat
irregular $\chi^2$ curves, we require that $\ddchiSq(t_n)\ge\epsilon$, 
where $1+\epsilon$ is the smallest representable number $>1$ on our machine
and we require that $| t_{n+1} - t_n | < 25$ years.
Additionally, if Newton's method appears to be diverging or failing to
converge, we find $t_{n+1}$ using a binary-search approach in which we sample
the midpoint of the area in which the minimum appears to lie. This collection
of restrictions effectively amounts intelligent gradient descent, which we
switch to when Newton's method cannot be performed. This fallback system is
rarely required, but does sometimes prevent oscillation and search failure.}

Once we have found $t^*$, we can locate $t^*_-$ and $t^*_+$. We use our
modified Newton's method with these new goals:
\begin{eqnarray}\displaystyle
	\chi^2(t^*_+) = \chi^2(t^*)+1 \nonumber \\
	\chi^2(t^*_-) = \chi^2(t^*)+1
\end{eqnarray}
This uncertainty region would be the true one-sigma uncertainty region in the
limit that the modified chi-squared were the standard linear-fitting
chi-squared. Because of the weighting function, in practice this criterion
over-estimates the one-sigma uncertainty.

We can perform root-finding on these equations using the following formula for
iteration:
\begin{equation}
	t_{n+1} = t_n - \frac{\chi^2(t_n) - \left( \chi^2(t^*)+1 \right)}{\dchiSq(t_n)}
\end{equation}

We begin our two new searches with a sensible initial estimate of the bounds
of the uncertainty region, based on our present knowledge of the $\chi^2$
curve. These searches operate under all of the constraints under which the
previously detailed search operated, and also converges when
$\norm{\chi^2(t_{n+1})-\chi^2(t_{n})} \leq \delta_\chi$ holds for two
consecutive iterations.

\begin{figure}

	\resizebox{\onewidth}{!}{\includegraphics{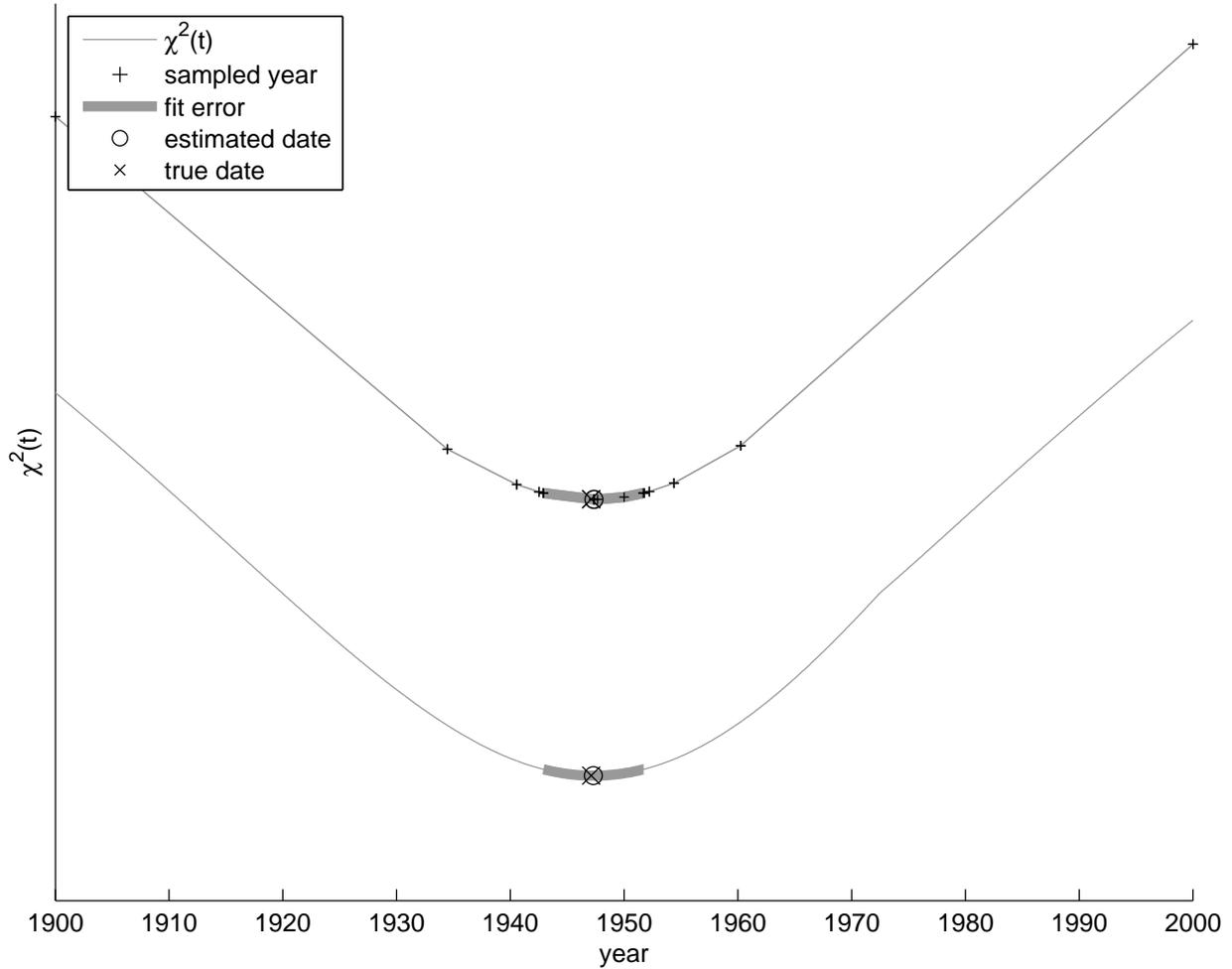}}

	\caption{
Evaluation of the accuracy of our search algorithm. The top is our modified
Newton's method with $\delta_\chi=10^{-4}$, and the bottom is brute force,
with an interval of $0.1$ years. The uncertainty regions are identical to
within $10^{-3}$ years, and the estimated dates are within $0.05$ years of
each other. Brute force took 1001 iterations to achieve this accuracy, while
Newton's method took only 23. For clarity's sake, the curves were vertically
separated, and the individual points that were sampled to construct the
brute-force curve are not displayed. \label{fig:searchMethodCompare}}
\end{figure}

\begin{figure}
\centering
\subfigure[Estimated $\chi^2$ curves of the science-quality images.]{
	\label{fig:chiSqResults1}
	\resizebox{\twowidthshort}{!}{\includegraphics{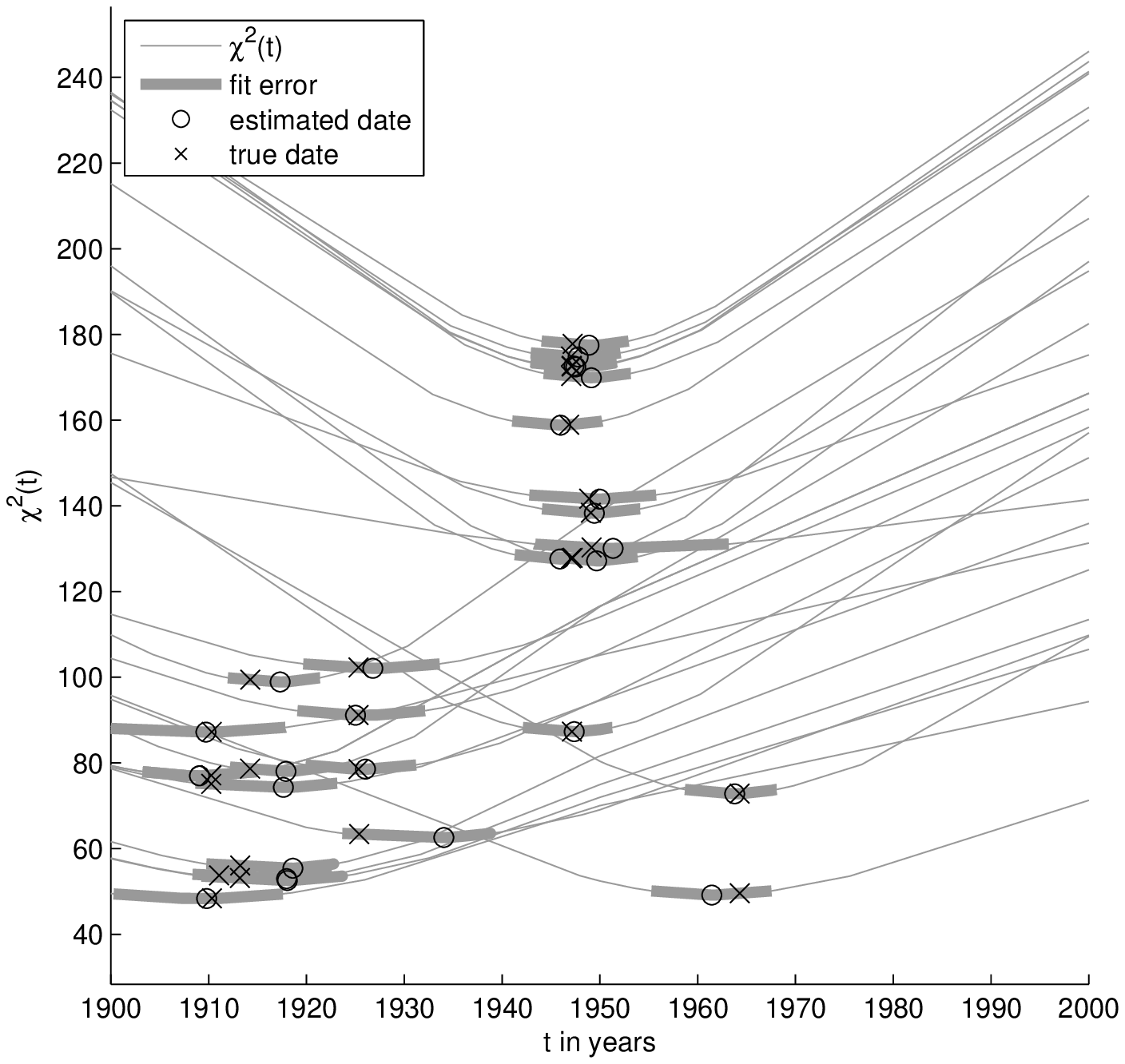}}
}
\subfigure[Estimated $\chi^2$ curves of the low-quality images.]{ 
	\label{fig:chiSqResults2}
	\resizebox{\twowidthshort}{!}{\includegraphics{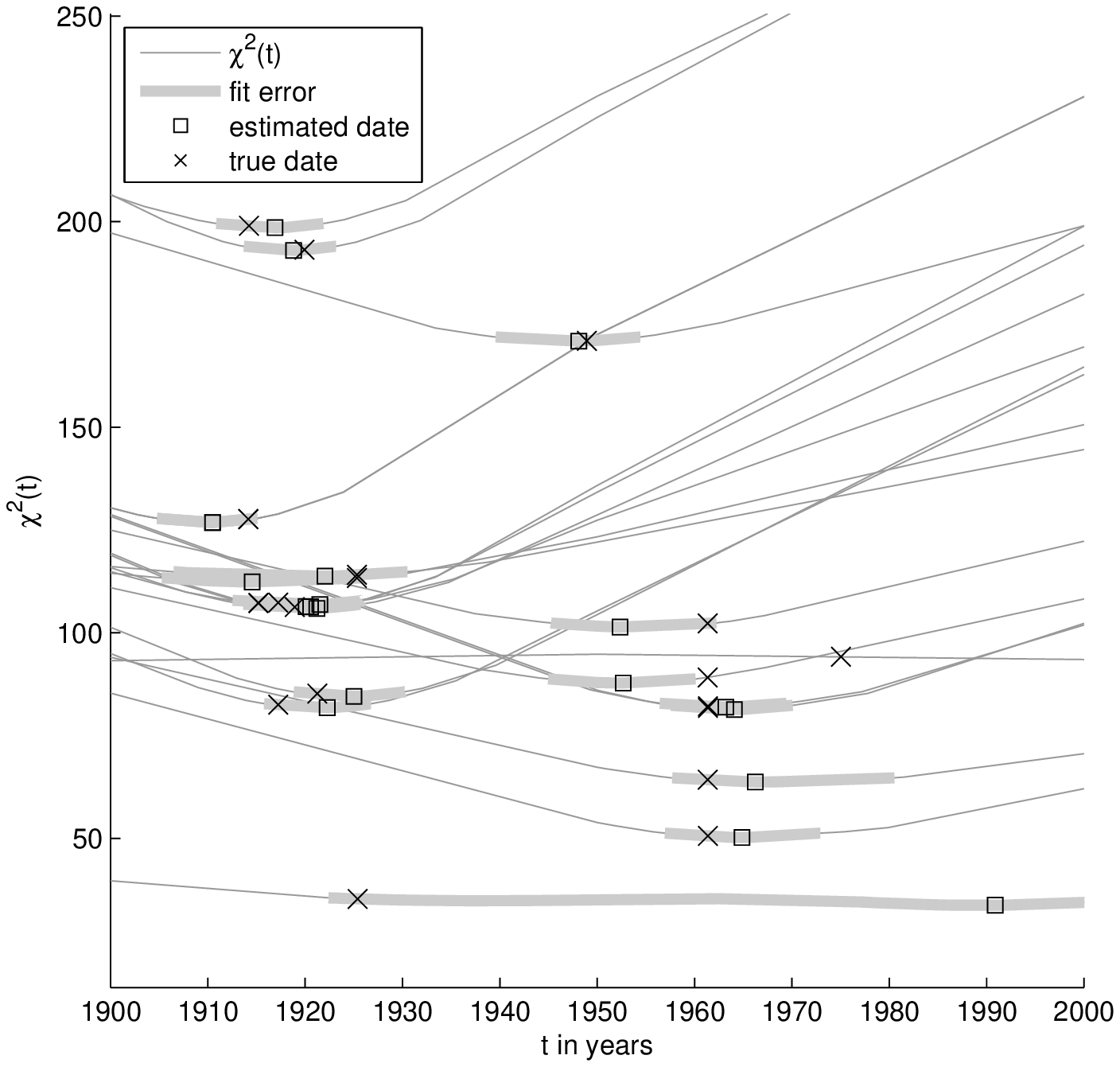}}
}

\caption{
The estimations of each image's $\chi^2$ curves generated by our modified
Newton's method, for both datasets. The vertical axes of these plots do not
show the constant contributions of image--catalog pairs whose separations are
always too large to be directly calculated, as including those would make
the curves excessively vertically separated, rendering these plots
incomprehensible.
\label{fig:chiSqResults}}
\end{figure}

Additionally, we can speed up the total search process by using the
transformed image points from the previous iteration to find the new
transformation for the next iteration. This heuristic means that as the search
converges on a final result, the amount of time required to query new years is
dramatically reduced. Also, it becomes easier to visualize the search
algorithm as a single bidirectional fitting process, in which we repeatedly
fit the image to the catalog and the catalog to the image until both fittings
have converged. Just as in the previous section, all transformations are
constructed using the initial positions of the points, so our final
transformation after searching the $\chi^2$ curve is still very numerically
stable. Figure~\ref{fig:searchMethodCompare} shows a comparison of this search
algorithm against a brute-force ``ground truth''.
Figure~\ref{fig:chiSqResults} shows the modified $\chi^2$ curves for each
image as they were estimated by this search algorithm.

The output of this process is a polynomial description of the image
astrometric WCS, a best-fit year value $t^*$, and an uncertainty region around
that value.

\subsection{Implementation notes}

\bd's two primary performance bottlenecks are calculating the distances
between image and catalog points and solving the weighted least-squares
problems. In both cases, we are able to use the properties of our weighting
function to construct approximate solutions that very effectively approximate
the true solution.

Our algorithm theoretically requires us to repeatedly calculate the distances
of all image--catalog pairs. However, due to the properties of the weighting
function as described in Section \ref{sec:objectiveSection}, we do not need to
know the distances of significantly separated pairs. Because the
transformation applied at each iteration tends to be very small, we can
generally assume that significantly separated pairs in one iteration will also
be significantly separated in the next iteration. This allow us to do one
initial calculation of all image--catalog distances, but in later iterations
only calculate the distances of image--catalog pairs that will likely cause a
change in the value of the objective function. This is a rough heuristic, so
we safeguard ourselves by manually recalculating all image--catalog distances
every 10 iterations, as well as whenever the IRLS begins to converge. This
dramatically speeds up our algorithm, while producing nearly identical results
to the na\"{\i}ve brute-force approach.

We use the aforementioned properties of our weight function to determine if an
image--catalog pair should be considered in the weighted least-squares
calculation. A highly separated pair always contributes a nearly-constant
value to the residuals, and therefore can be safely ignored. This gives us a
slight performance boost.

We require an additional threshold for the difference between the optimal
function from one iteration to the next, which determines when our IRLS
operation has converged. We use the very conservative value of $10^{-4}$ as
the maximum amount that the $\chi^2$ score of the final IRLS iteration is
allowed to change from those of the previous two IRLS iterations.

\section{Results}

\begin{figure}
\centering
\subfigure[Performance on the science-quality images.]{ 
\label{fig:performance1}
\resizebox{\twowidthshort}{!}{\includegraphics{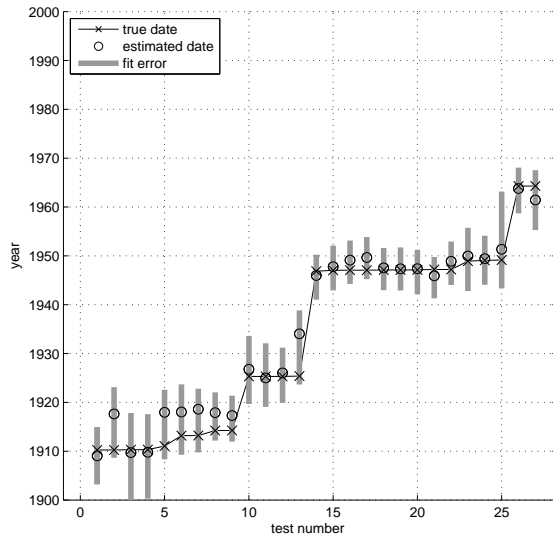}}
}
\subfigure[Performance on the low-quality images.]{
\label{fig:performance2}
\resizebox{\twowidthshort}{!}{\includegraphics{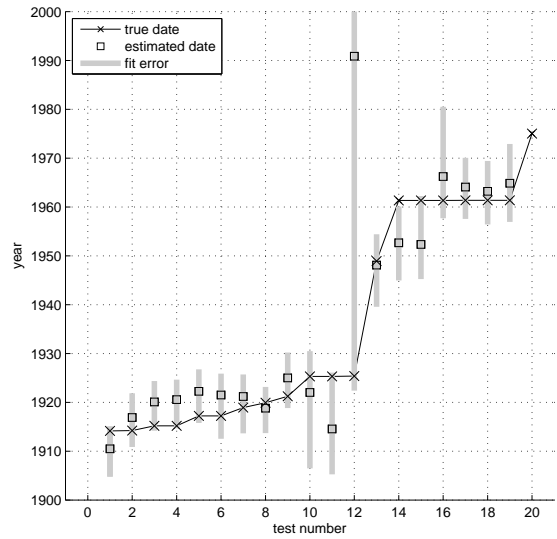}}
}

	\caption{
An informative visualization of performance for both datasets. The x-axis is
an arbitrary index denoting the test image (images were sorted by true year),
and the y-axis shows each image's true year, estimated year, and uncertainty
region. The last image in figure~\ref{fig:performance2} is not shown, as
it is estimated as originating before $1900$.
	\label{fig:performance}}
\end{figure}

Harvard's interface for accessing its scanned plates of M44 made it difficult
to obtain more than $3000$ by $3000$ pixel subsets of the images, though the
entire plates are significantly larger. The interface for downloading the
images did not provide an obvious mechanism for selecting the same $3000$ by
$3000$ pixel subsets of each image, which means that such selection was done
by hand, and is therefore not very accurate. Many of the plates suffer from
the many sources of noise typical of historical imagery: Some are
multiple-exposures, badly out of focus, badly saturated, or cracked, and some
contain handwritten labels, digital scanning artifacts, and bad trailing. For
the sake of fairly assessing \bd's performance, we split the images into two
sets; \numcleantests\ ``science-quality'' images and \numnoisytests\
``low-quality'' (see Figure~\ref{fig:imageExamples} for examples). For our
convenience, we used the JPEG versions of the images, which probably
introduces some minor noise in the form of compression artifacts. Harvard
graciously provided ground-truth dates for each image, which presumably were
taken from logs or from writing on the plates. We take these dates to be true.
Though the ground-truth dates range from 1910 to 1975, the dates are not
uniformly distributed. See the distribution of images along the y-axis of
figure~\ref{fig:performance} for a demonstration of this clustering.

The tests were performed using the modified Newton's method, with affine
distortions and an accuracy threshold $\delta_\chi$ of $10^{-4}$ year. Tests
were done on a 2007 Macbook with a 2GHz Intel Core 2 Duo processor, and 2GB of
RAM. Median runtime for estimating each image's date was $\sim 2.6$ seconds
per image, after source extraction and \an's initial calibration. Results are
shown in Table~\ref{table:performance}.

\begin{table}[!h]
	\begin{center}
	  \begin{tabular}{| l || c | c | c |}
	    \hline
		  & mean year error (bias) & median absolute error & fraction within uncertainty \\ \hline \hline
		science-quality & $ 1.68 $ & $ 1.29 $ & $ 27/27 $ \\ \hline
		low-quality & $ -5.28 $ & $ 4.00 $ & $ 17/20 $ \\ \hline
		\end{tabular}
	\end{center}
	\caption{Accuracy of estimated dates for the two subsets of data. 
	See Figures \ref{fig:performance} and \ref{fig:errorDist} for a
	more detailed visualization of the results.
	}
	\label{table:performance}
\end{table}

Though the results shown were generated by fitting an affine (linear)
transformation in image coordinates, we experimented with increasing the order
of the polynomial warp being fitted. Results were very similar to those
obtained using affine transformations, although the median absolute error
for science-quality images dropped to $1.01$ years for second-order
transformation, and to $0.99$ years for third-order transformations. Median
absolute error for low-quality images also decreased slightly as the order was
increased, as did the bias for science-quality images.

Additionally, we tested \bd\ on the five \USNOB\ source images of M44 that we
were able to retrieve from the US Naval Observatory Precision Measuring
Machine Data Archive, and on the Sloan Digital Sky Survey \citep{york00} image
of M44. The dates of the \USNOB\ source imagery were estimated accurately (all
within six years of the true dates, and all within the uncertainties), which
is as good as we would expect performance to be given the relatively low
resolution ($3.2\,\arcsec\,\pix^{-1}$) of the source imagery that we were able
to obtain. The SDSS image was estimated to have been taken in late 2004, and
was actually taken in 2006.

We assembled an alternate test-bed of ten amateur images of M44 that we were
able to find online. The websites on which we found most of our imagery did
not explicitly note the date at which the image was taken, so we were forced
to use the ``date'' tag in each image's EXIF meta-data as the ground truth.
For many of the images, \bd\ provided accurate dates and uncertainties that
are consistent with our previous findings: all estimated dates lay within the
our uncertainty bounds, accuracy generally depended on the resolution and
quality of each image, and most estimated dates (for all sufficiently
high-resolution images) were within a few years from the true dates. Our
ground-truth dates are, unfortunately, very unreliable, as the EXIF data may
simply reflect the date at which an image was digitized or modified, rather
than the date at which it was imaged. Though this means that we are not able
to truly vet \bd's performance for these amateur images, this issue also
highlights the utility of this system: the dates of origin of these images are
effectively lost, but can be re-estimated.

\section{Discussion}

We have shown that our \bd\ system can successfully attach time meta-data to
historical imaging data. The system runs in seconds on standard inexpensive
consumer computer equipment; it does not require large investments of time or
money to vet or create time meta-data for large collections of astronomical
imaging.

The performance of \bd\ will depend on the properties of the input image, and
on the properties of the catalog information known about the region of the sky
that is being imaged. We can phrase this as two questions: ``What is the
information content in an image?'', and ``What is the information content in a
catalog star?''

\begin{figure}
\centering
\subfigure[\examplecaptionA]{
	\label{fig:imageExample1}	
\resizebox{\threewidthshort}{!}{\includegraphics{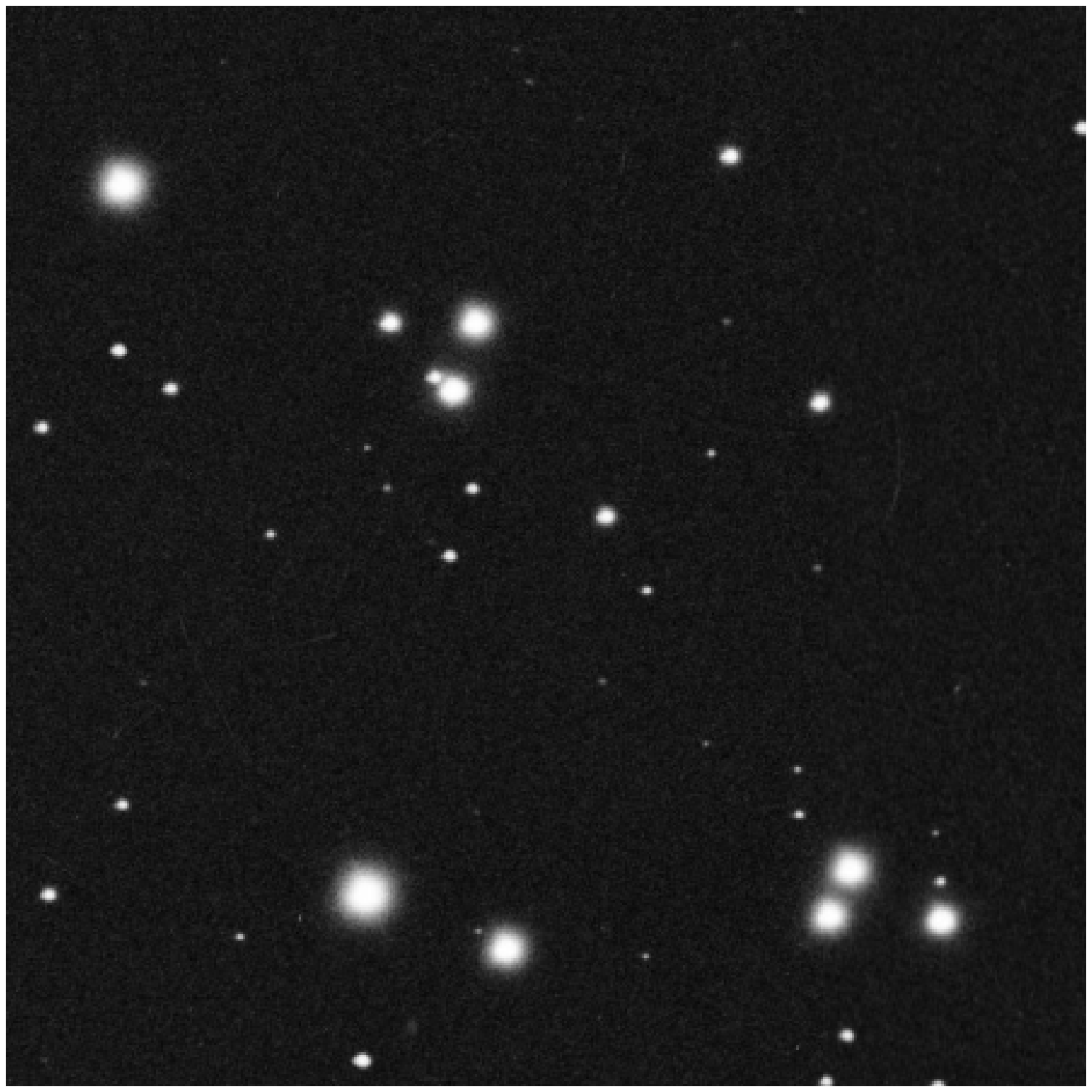}}
}
\subfigure[\examplecaptionB]{
	\label{fig:imageExample2}
	\resizebox{\threewidthshort}{!}{\includegraphics{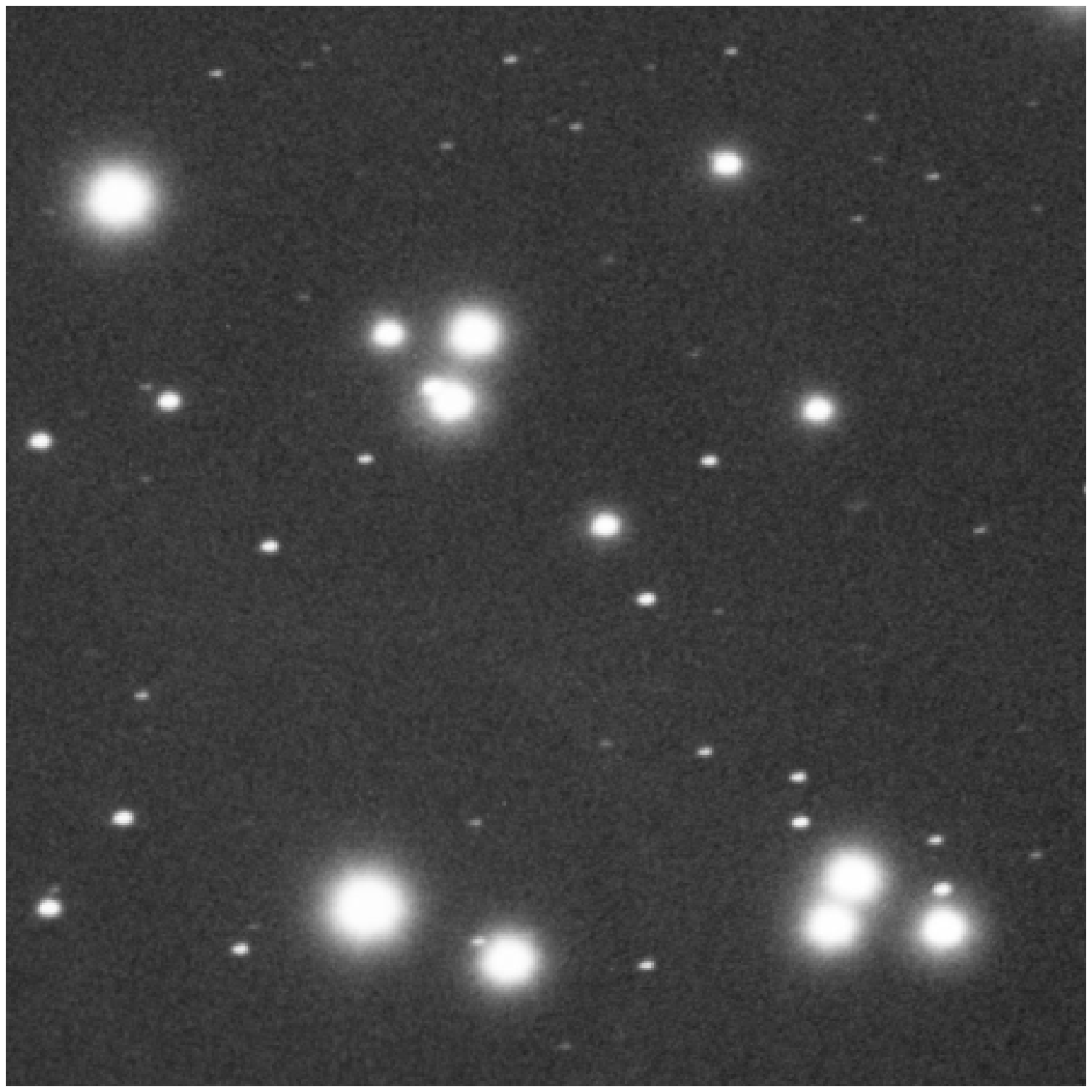}}
}
\subfigure[\examplecaptionC]{
	\label{fig:imageExample3}
	\resizebox{\threewidthshort}{!}{\includegraphics{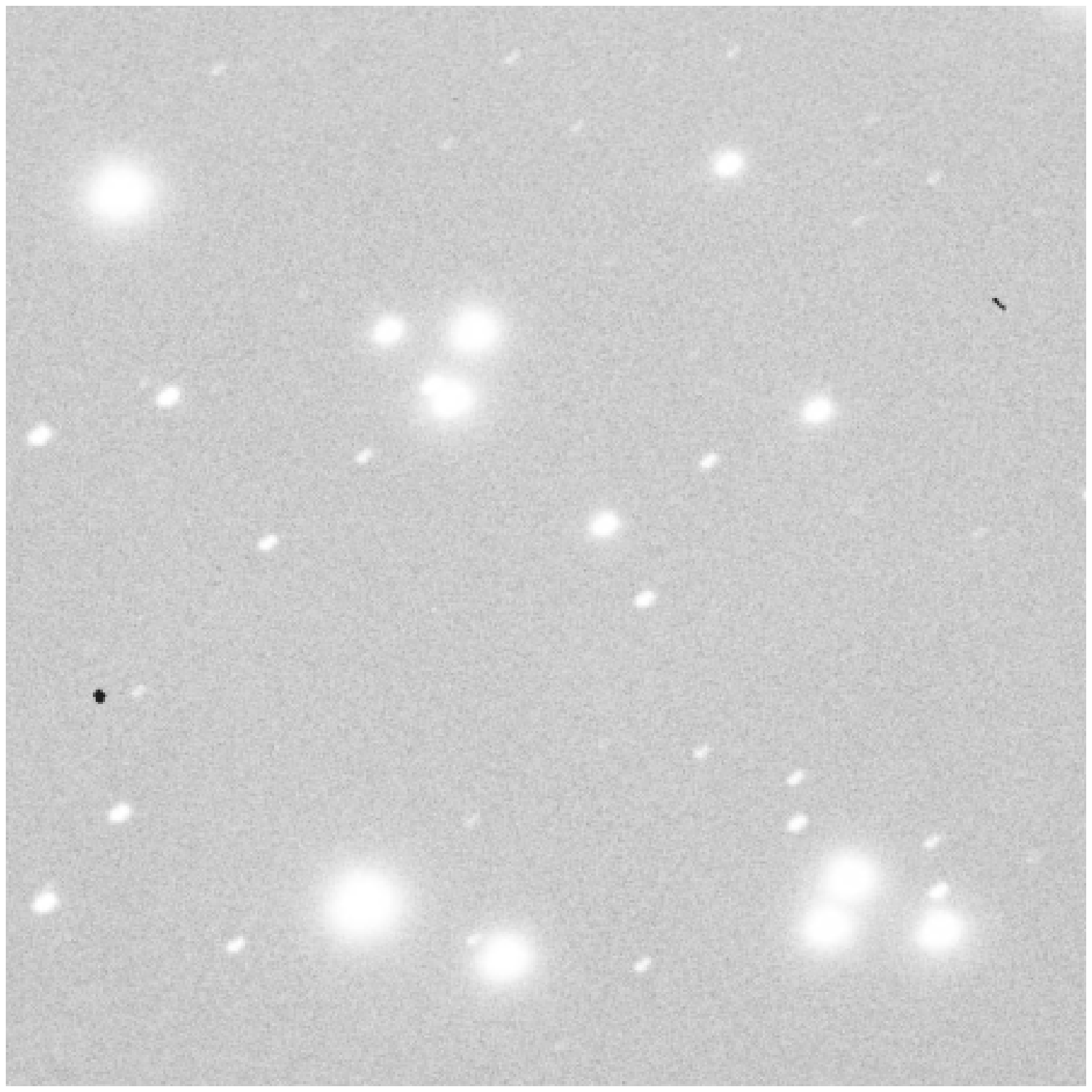}}
}
\subfigure[\examplecaptionD]{
	\label{fig:imageExample4}	
	\resizebox{\threewidthshort}{!}{\includegraphics{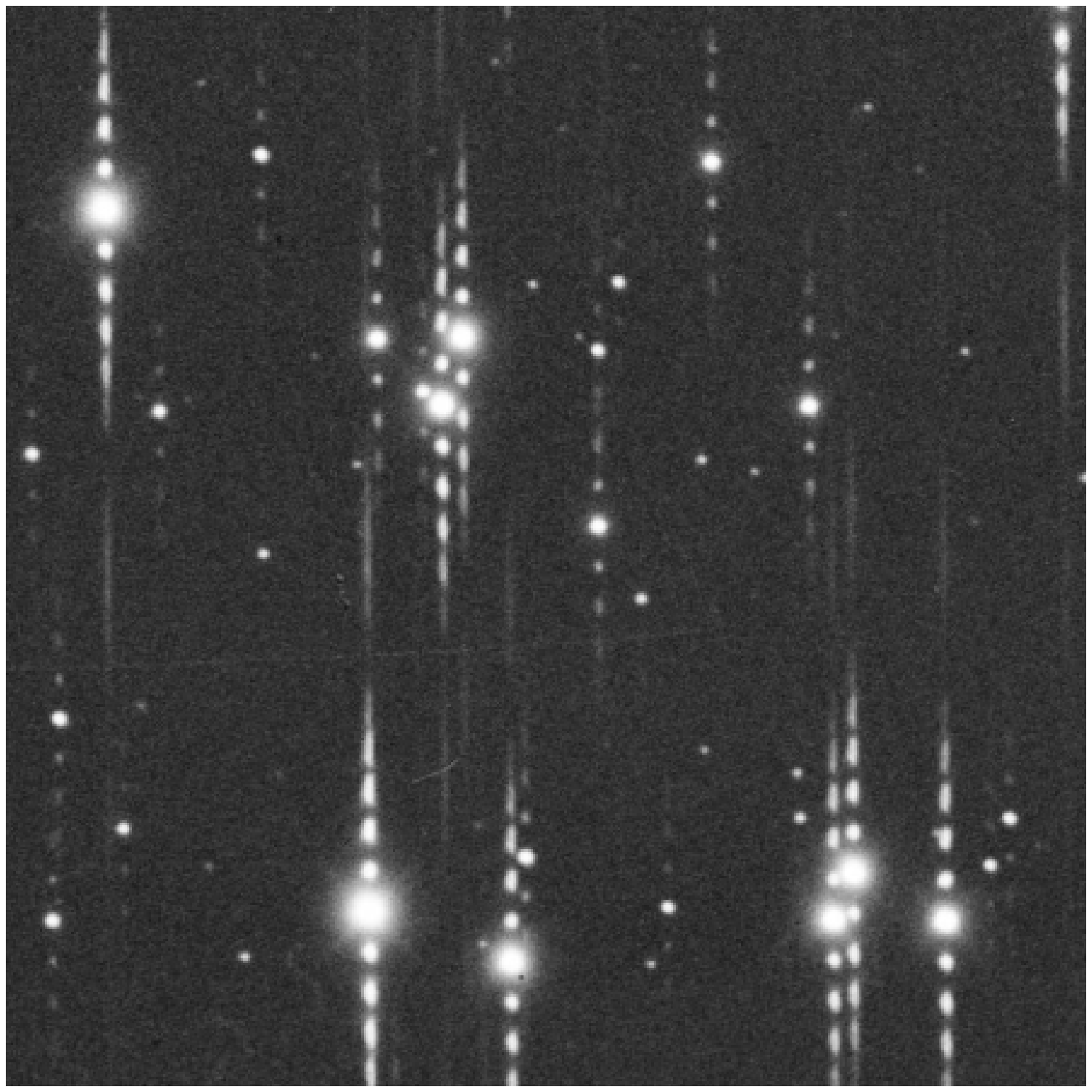}}
}
\subfigure[\examplecaptionE]{
	\label{fig:imageExample5}	
	\resizebox{\threewidthshort}{!}{\includegraphics{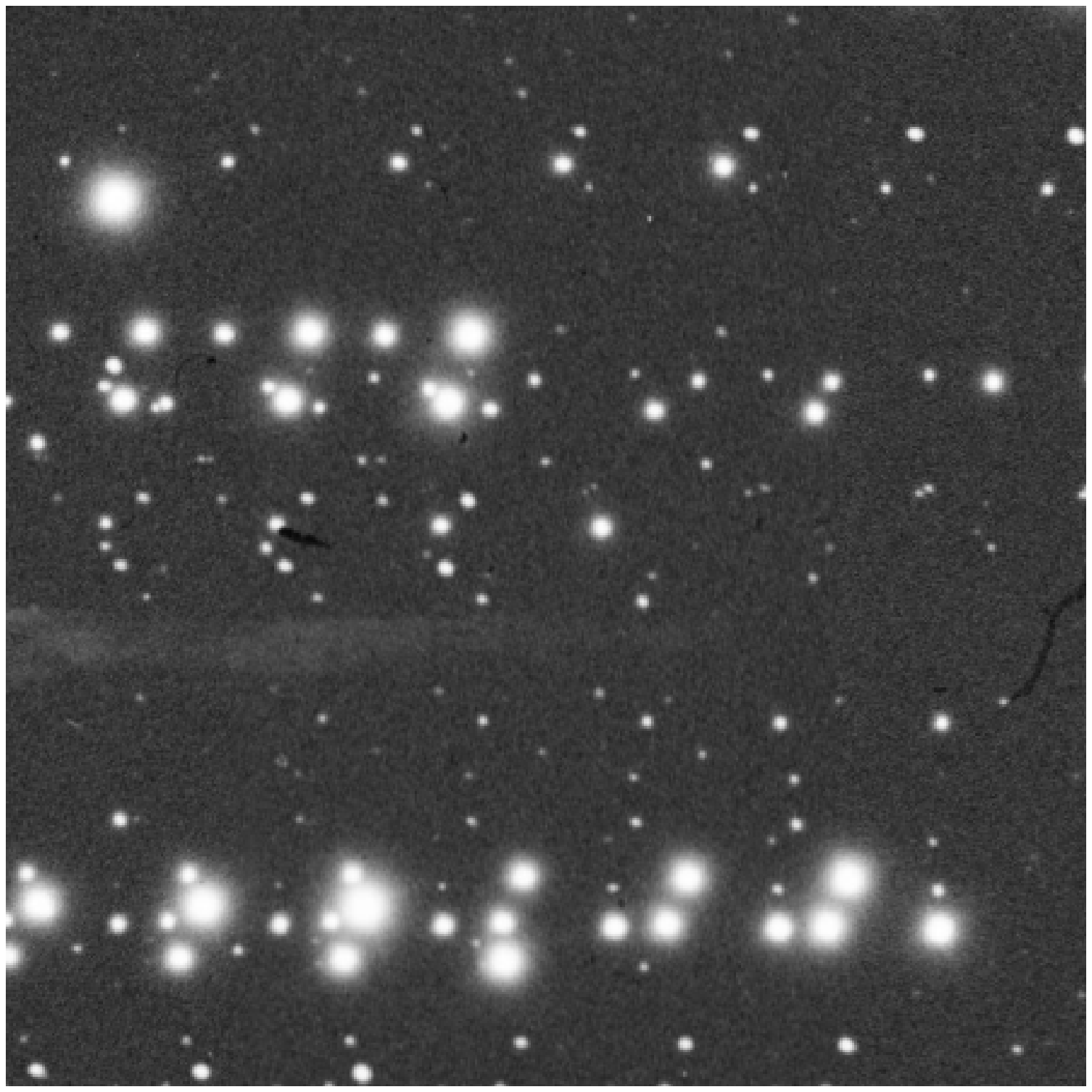}}
}
\subfigure[\examplecaptionF]{
	\label{fig:imageExample6}	
	\resizebox{\threewidthshort}{!}{\includegraphics{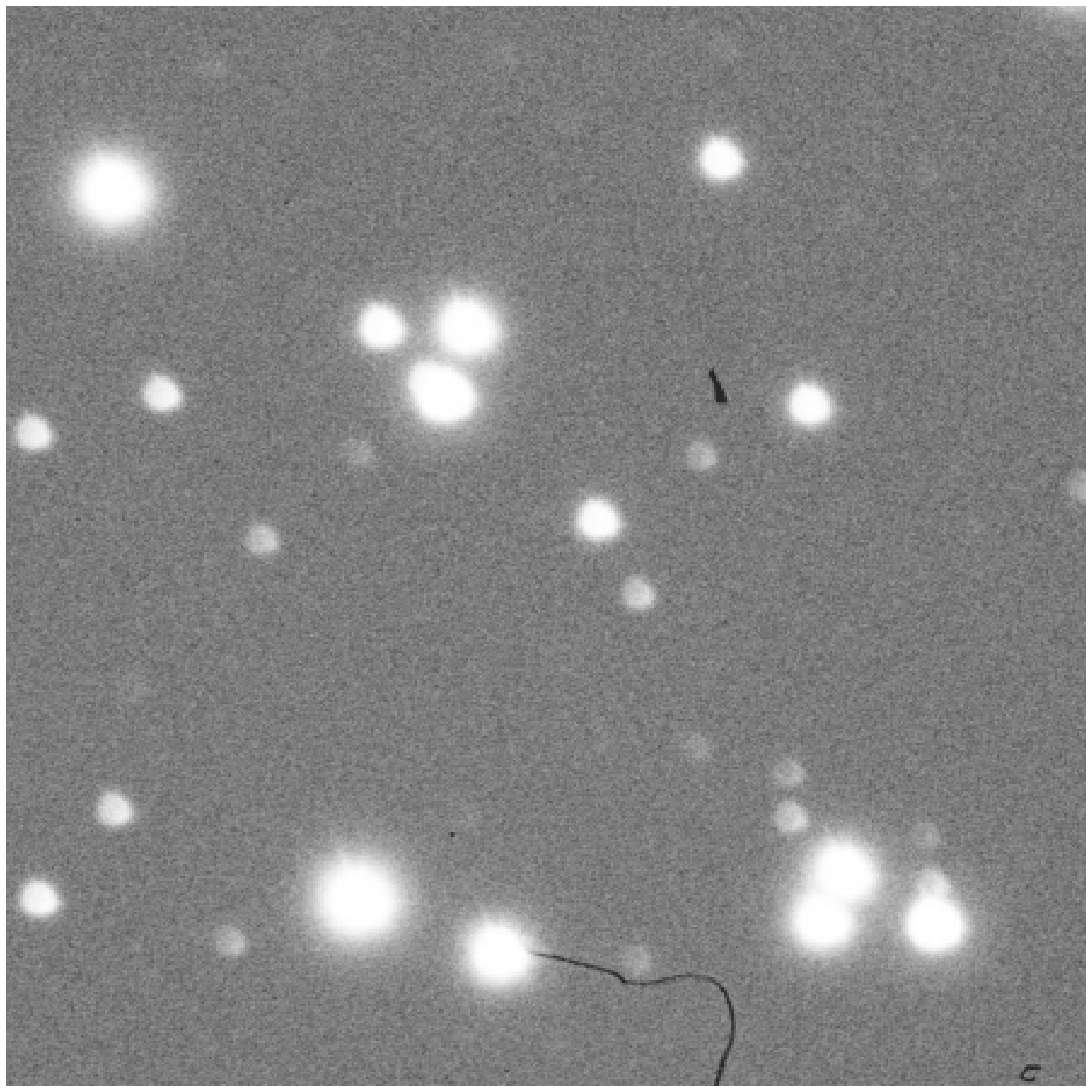}}
}

	\caption{
A series of $900$ by $900$ pixel subsets of our $3000$ by $3000$ pixel images.
The top three images are from our set of science-quality images, and the the
bottom three images are from our set of low-quality images. The captions are
of the form ``true date / estimated date''. All but
figure~\ref{fig:imageExample6} have estimated dates that lie within the
uncertainty region.
	\label{fig:imageExamples}}
\end{figure}

\begin{figure}

	\centering
	\subfigure[Performance versus resolution.]{
		\label{fig:performanceVs1}
		\resizebox{\twowidthshort}{!}{\includegraphics{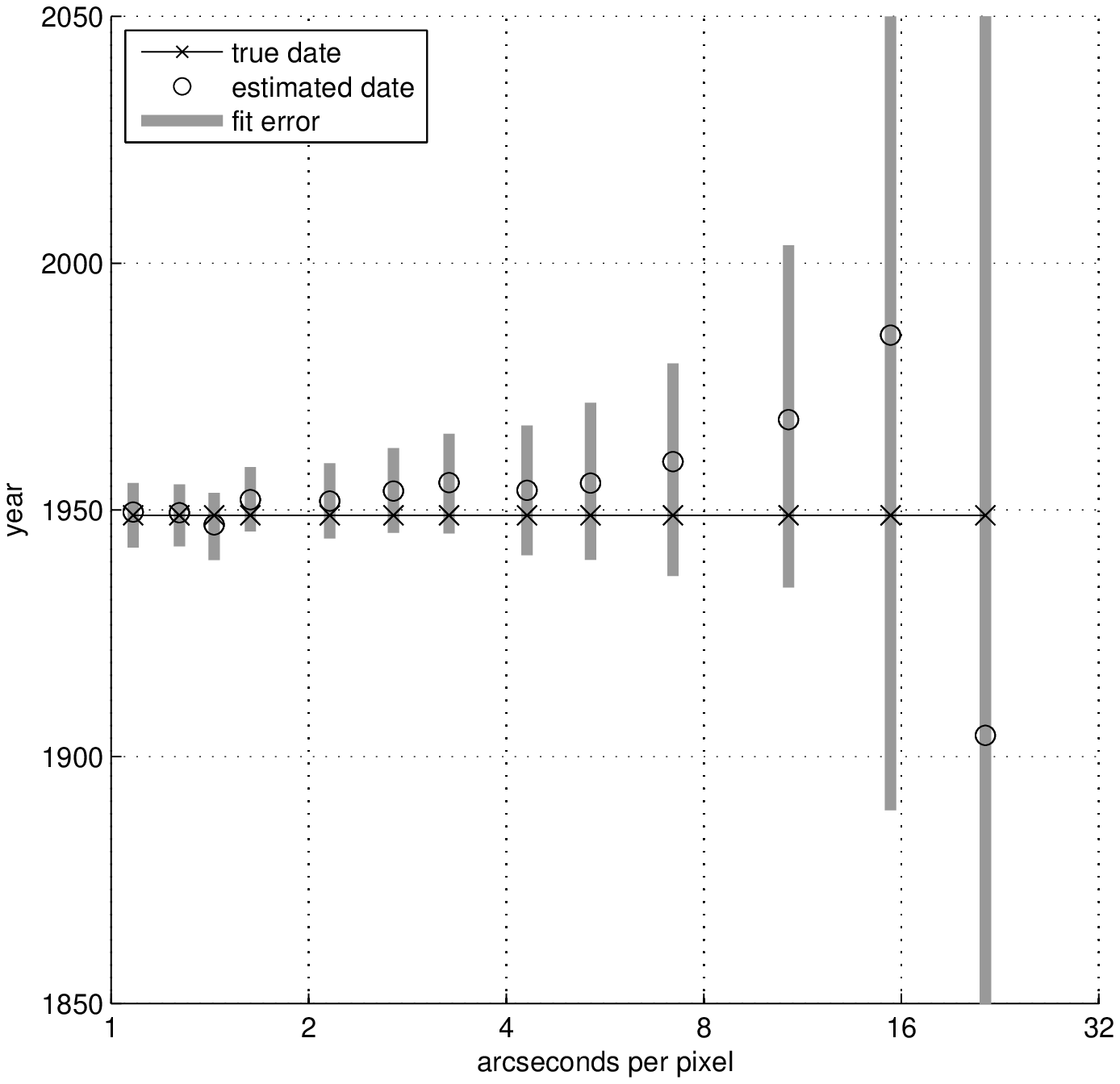}}
	}
	\subfigure[Performance versus number of image stars.]{
		\label{fig:performanceVs2}
		\resizebox{\twowidthshort}{!}{\includegraphics{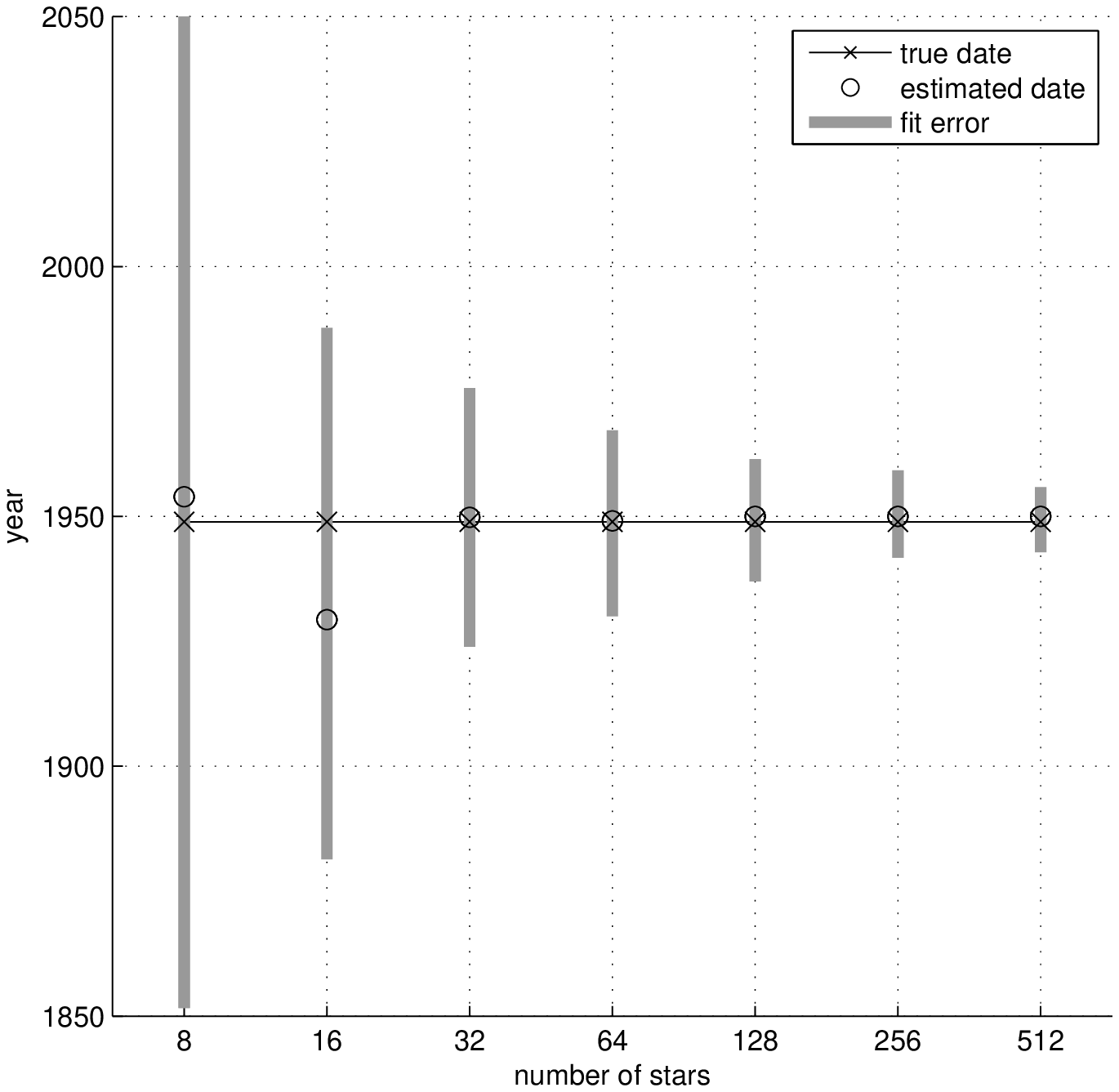}}
	}

	\caption{
Plots showing performance relative the resolution of the input image, and to
the number of stars the input image contains. Figure~\ref{fig:performanceVs1}
was produced by repeatedly downsampling the input image, while
Figure~\ref{fig:performanceVs2} was produced by repeatedly cropping out the
borders of the input image. Note that in downsampling the image, some smaller
stars stop being detectable by our source-extraction algorithm, so the number
of stars in the image decreases with the resolution of the image.
	\label{fig:performanceVs}}
\end{figure}

In an attempt to empirically assess the information content of an image, we
ran a simple experiment in which we varied the resolution of an input images
and the number of stars contained in an input image (by downsampling and
cropping the image, respectively). The results are shown in
Figure~\ref{fig:performanceVs}, where we see that performance depends heavily
on an image containing a large number of stars imaged at high resolution. We
also explored the effects of different kinds of imaging defects. Our objective
function is designed to be robust to false sources, and as such, \bd\ performs
very well on images with multiple exposures. Our experiments suggests that
saturation, large PSF due to poor focus or trailing, and short exposure time
(low sensitivity) most negatively effect \bd's performance. See
Figure~\ref{fig:imageExamples} for examples of our accuracy in the face of
different kinds of imaging defects. Trailing and saturation can effectively be
thought of as decreasing the resolution of our input image (by decreasing our
ability to accurately centroid stars), and shallow imaging is effectively
equivalent to dropping dim stars out of the image; these are the two trends
demonstrated in Figure~\ref{fig:performanceVs}. Once again, \bd's accuracy
appears to depend on an image containing many well-imaged (high resolution)
stars.

The information in a single catalog star (provided that it has been
detected in the input image, in the limit that our procedure is
equivalent to least-square fitting) is proportional to that star's
contribution to the second derivative of $\chi^2$ with respect to
date. This contribution is approximately the square of the magnitude
of the star's proper motion, divided by the square of the uncertainty,
that is, the square of the signal-to-noise at which the proper motion
is detected (where the ``noise'' in this case is the combined
uncertainty from the catalog and the image as in
equation~\ref{eq:uncertainty}).  We expect \bd's performance on an image to
scale roughly with the sum of the squares of the detected catalog
stars' proper motion signal-to-noise ratios.  Imagery unlike the
imagery analyzed here ought to obtain date calibration with
uncertainty that goes down as the sum of the detected catalog stars'
proper motion signal-to-noise ratios goes up.

Increasing the polynomial order of the transformation on the image plane
produces slightly more accurate date estimates, presumably because the input
images do have distortions that are represented reasonably by these functions.
We are reluctant to advocate unnecessarily large polynomial orders,
as---theoretically---the more freedom we give the transformation model, the
more irregular our resulting $\chi^2$ curves may become. That being said, we
have not seen any evidence that reinforces such a concern. In principle, even
more accurate results could be obtained without increasing the number of
degrees of freedom in the fit by employing a physical camera model that
represents known distortions in the particular camera used to take the data.

For each image in our dataset, we performed an experiment to determine the
range of initial dates for which our search algorithm is robust. We discovered
each image had a range of at least $300$ years (and on average, $620$ years)
roughly centered around the true year from which the search could be
initialized without the final result being affected. If we ignore our
precaution of using gradient descent when the second derivative of the
$\chi^2$ curve is non-positive, this range is significantly smaller (on
average, about $55$ years). This finding highlights the importance of the
modifications we make to Newton's method in constructing our search algorithm,
and also suggests that the coarse grid of queries with which we initialize our
search is unnecessary---a single initial query in the correct century would
have been more than sufficient.

\bd\ largely ignores one very important source of information, namely the
brightnesses of the image and catalog stars. This data is used in our
band-pass estimation step, but is then largely ignored. In future versions of
\an\ we plan to utilize this data in a number of ways. We eventually hope to
simultaneously estimate all parameters of a given image, including the image's
WCS, its date of origin, and its band-pass. Estimating all of these
simultaneously should means that brightness information is implicitly used in
our estimation of the location and date, and should improve our results
accordingly. This would also solve our current conundrum regarding this
process, which is that image--catalog correspondences are required for
band-pass estimation, while band-pass estimation is required for finding
image--catalog correspondences.

Analysis of the brightness of image stars may be able to play a profound role
in date estimation if we consider the subset with periodic variability. Given
an image containing $k$ stars with different periods, and given sufficient
information concerning the periodic variations in their brightnesses, we
should be able to constrain the date of origin of the image to one of a set of
time intervals in which those $k$ stars are at whatever particular point in
their periods (to within photometric precision). Given the set of intervals
constrained by the periodic variations, we can use the range of dates
determined by \bd\ (that is, from the proper motions of the stars) to select a
potentially very narrow time interval in which the image must have originated.
In principle, it may even be possible to determine the date of origin of an
image solely though periodic brightness, though that would require very good
measurements of the periods of the catalog stars and of the brightnesses of
the image stars.

\bd's value, on the most superficial level, is clear: this system could be
used to recover lost meta-data (at low precision) for historical and amateur
data that have been archived poorly or not at all. Now that large scanning
projects are underway at photographic archives and the web is providing new
opportunities for file sharing among amateurs and professionals, we need
systems that automatically vet and provide meta-data for data of unknown
provenance.

Regardless of whether or not imagery already contains reliable date meta-data,
the techniques described in \bd\ may have deep-seated implications for the
calibration of all imagery not taken at the year $\epoch$. The fact that the
date can be well estimated from input images demonstrates both that the images
contain important information about stellar motions, \emph{and} that
astrometric calibration is hampered when calibration is performed with a
catalog projected to an epoch far from the date of the image. A system that is
time sensitive, such as \bd, will plausibly provide the best astrometric
calibration possible for arbitrary imaging.

What may be \bd's most important consequence is an inversion of the system,
in which we attempt to use imagery to re-estimate the proper motions of
catalog stars. The most straightforward approach to this would be to ``cheat''
and use the ground-truth dates of all input imagery. One could repeatedly:
calibrate each image using the catalog wound to that image's date-of-origin,
re-estimate the proper motions of the catalog stars, and re-wind the catalog
using those new proper motions. This could be thought of as performing
expectation-maximization on the proper motions of the catalog. Of course, an
ideal system would be robust to some (or all) input imagery not having
ground-truth ages. We could estimate the date-of-origin of all unlabeled
imagery, and use these estimates (and their uncertainties) in our
expectation-maximization. Labeled and unlabeled data could be treated
equivalently, except that labeled data would have much less uncertainty
associated with it. This system would then become a two-way street, in which
we do not just reposition images relative to the sky, but also reposition the
sky relative to the images, and dynamically develop a consensus between the
two. The future \an\ ``catalog'' would not have to be a static entity, but
would instead be a consensus of all available imagery---using the \USNOB\ as a
static ``prior.'' \bd\ takes us one step closer to this grand long-term hope
for \an\ becoming an always-changing database of everything we know about the
sky, by allowing \emph{time} to become one more dimension of our data.

\acknowledgments We would like to acknowledge generous assistance from Mike
Blanton, Rob Fergus, Yann LeCun, Brett Mensh, Keir Mierle, and Dave Monet. We
thank the USNO-B and DASCH teams for providing the data used for this study.
This project made use of the NASA Astrophysics Data System, the US Naval
Observatory Precision Measuring Machine Data Archive, and data and code from
the \an\ project.

\clearpage

\begin{figure}
\centering
\subfigure[Error in date estimation.]{
	\label{fig:errorDist1}
	\resizebox{\twowidthshort}{!}{\includegraphics{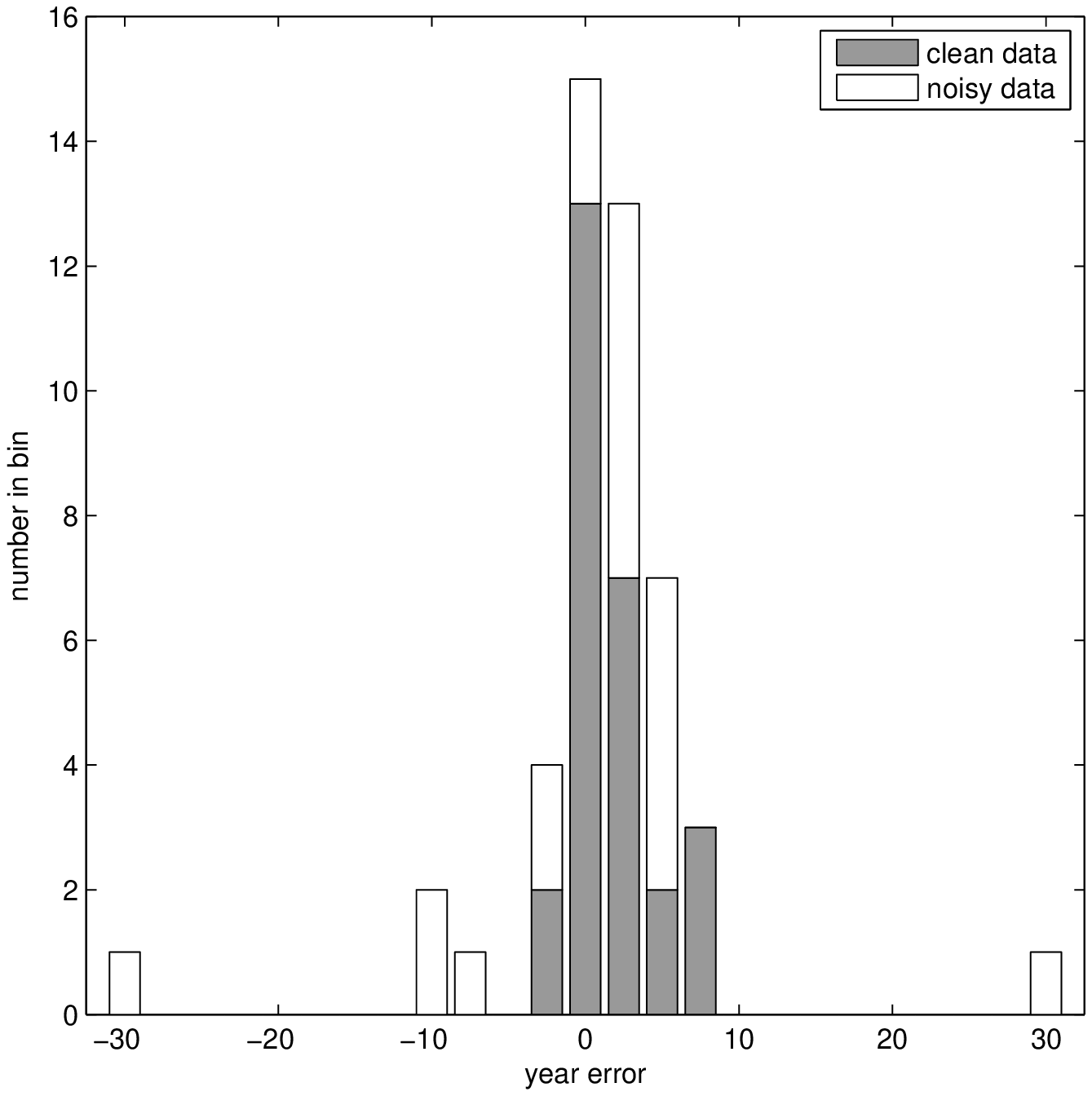}}
}
\subfigure[Error relative to uncertainty.]{
	\label{fig:errorDist2}
	\resizebox{\twowidthshort}{!}{\includegraphics{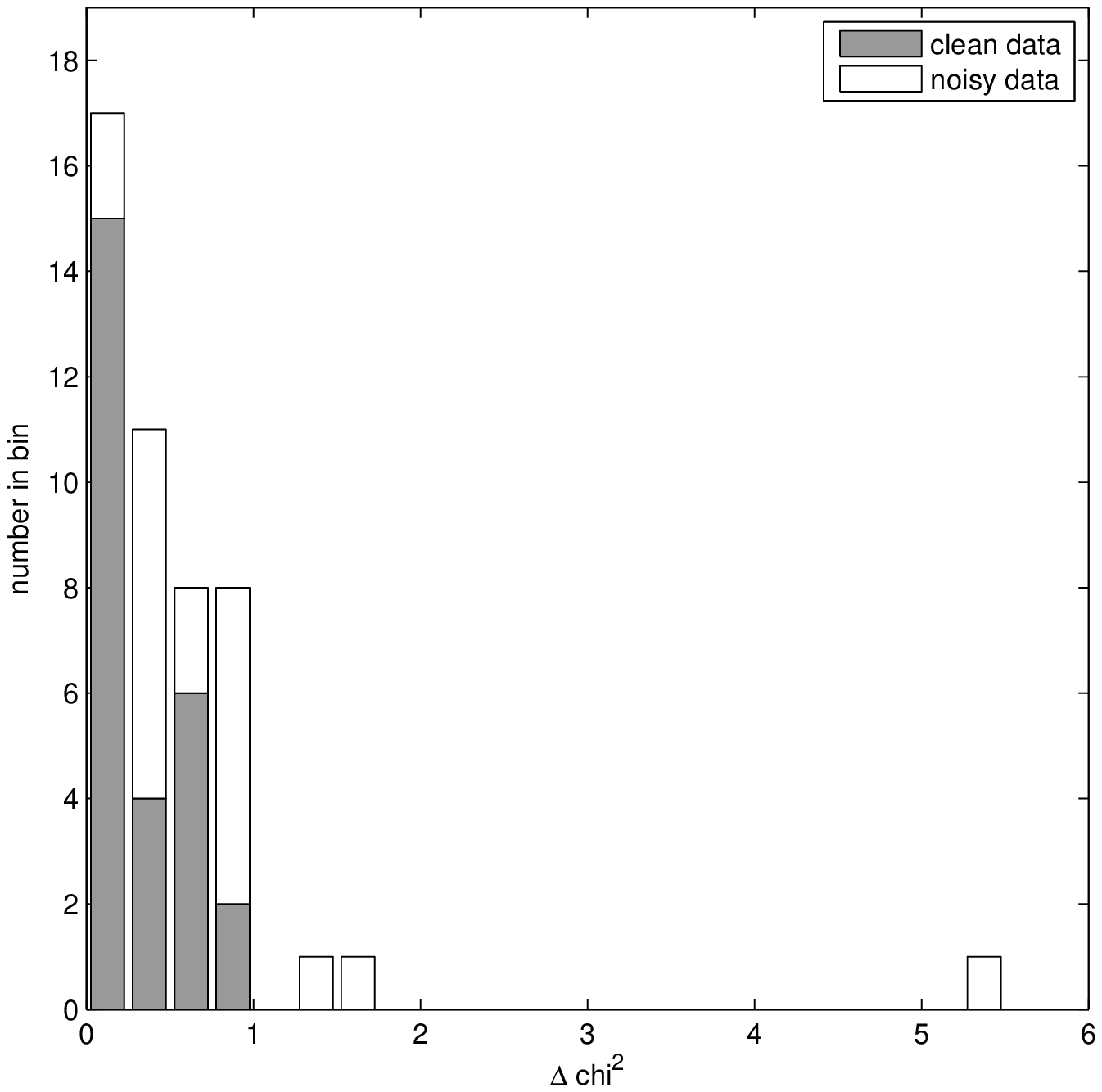}}
}

	\caption{
Histograms of the errors in date estimation, for both datasets.
Figure~\ref{fig:errorDist1} shows the difference between our estimated dates
and the true dates. Figure~\ref{fig:errorDist2} shows those errors relative to
the widths of the uncertainties; effectively, the differences between the
$\chi^2$ scores of the true years and the $\chi^2$ scores of the estimated
years. A relative error of less than $1$ indicates that the true year lies
within the uncertainty. The two outliers in figure~\ref{fig:errorDist1} are
actually significantly worse than they appear: $+65$ and $-175$ years.
	\label{fig:errorDist}}
\end{figure}

\end{document}